\documentclass[12pt]{article}
\usepackage{latexsym}
\usepackage{amsfonts}
\usepackage{amsmath}
\usepackage{psfrag}
\usepackage{graphicx}
\usepackage{amssymb} 
\usepackage{bbold}
\usepackage{pstricks,pst-node,pst-tree}
\usepackage{array}
\usepackage{rotating}

\begin{document}

\def\ovr{\overline r}

\def\ttd{\tt d}
\def\ttg{\tt g}
\def\ttG{\tt G}
\def\bttg{\mbox{\boldmath${\ttg}$}}

\def\wt{\widetilde}
\def\wth{\wt h}
\def\wti{\wt i}
\def\wtj{\wt j}
\def\wtx{\wt x}
\def\wty{\wt y}

\def\wtalpha{\widetilde\alpha}
\def\gam{\gamma}
\def\Gam{\Gamma}
\def\bXi{\mbox{\boldmath${\Xi}$}}
\def\lam{\lambda}
\def\Lam{\Lambda}
\def\bmu{\mbox{\boldmath${\mu}$}}
\def\bnu{\mbox{\boldmath${\nu}$}}
\def\bphi{\mbox{\boldmath${\phi}$}}
\def\vphi{\varphi}
\def\bvphi{\mbox{\boldmath${\vphi}$}}
\def\bpsi{\mbox{\boldmath${\psi}$}}
\def\btau{\mbox{\boldmath${\tau}$}}
\def\om{\omega}
\def\Om{\Omega}
\def\bom{{\mbox{\boldmath$\om$}}}
\def\bOm{\mbox{\boldmath${\Om}$}}

\def\omu{\overline\mu}
\def\obmu{\overline\bmu}
\def\ovphi{\overline\vphi}
\def\obvphi{\overline\bvphi}
\def\obpsi{\overline\bpsi}

\def\oom{\overline\omega}
\def\oOm{\overline\Omega}
\def\sbom{\mbox{\sl${\bom}$}}
\def\obom{\overline\bom}
\def\0bom{{\bom}^0}
\def\0obom{{\obom}^0}
\def\nbom{{\bom}_n}
\def\0nbom{{\bom}_{n,0}}
\def\n*bom{{\bom}^*_{(n)}}

\def\utheta{\underline\theta}

\def\wtbom{\widetilde\bom}
\def\whbom{\widehat\bom}
\def\oom{\overline\om}
\def\wtom{\widetilde\om}

\def\obOm{\overline\bOm}
\def\whbOm{\widehat\bOm}
\def\wtbOm{\widetilde\bOm}

\def\oomega{\overline\omega}
\def\oUpsilon{\overline\Upsilon}
\def\wtomega{\widetilde\omega}
\def\wtheta{\widetilde\theta}

\def\fB{\mathfrak B}
\def\fG{\mathfrak G}
\def\fW{\mathfrak W}

\def\rd{\rm d}
\def\rr{\rm r}
\def\rx{\rm x}
\def\ry{\rm y}
\def\rtr{\rm{tr}}

\def\bcH{\mbox{\boldmath${\cH}$}}

\def\cl{\centerline} 

\def\cA{\mathcal A} 
\def\cB{\mathcal B}
\def\cC{\mathcal C}
\def\cE{\mathcal E} 
\def\cF{\mathcal F} 
\def\cH{\mathcal H} 
\def\cK{\mathcal K} 
\def\cL{\mathcal L} 
\def\cT{\mathcal T}
\def\cV{\mathcal V} 
\def\cW{\mathcal W}
\def\ocW{\overline\cW}

\def\bbC{\mathbb C}  
\def\bbE{\mathbb E}    
\def\bbP{\mathbb P}
\def\bbQ{\mathbb Q}
\def\bbR{\mathbb R}
\def\bbS{\mathbb S}
\def\bbT{\mathbb T}
\def\bbZ{\mathbb Z}

\def\obbP{\overline\bbP}

\def\oJ{\overline J}
\def\oP{\overline P}

\def\ba{\mathbf a}
\def\bA{\mathbf A}
\def\bB{\mathbf B}
\def\bF{\mathbf F}
\def\bH{\mathbf H}
\def\bI{\mathbf I}
\def\bN{\mathbf N}
\def\bR{\mathbf R}
\def\bU{\mathbf U}
\def\bX{\mathbf X}
\def\bx{\mathbf x}
\def\oB{\overline B}
\def\obx{\overline{\mathbf x}}
\def\ox{\overline x}
\def\wtbx{\widetilde\bx}
\def\uk{\underline k}
\def\un{\underline n}
\def\ux{\underline x}
\def\uy{\underline y}
\def\wtux{\widetilde\ux}
\def\uX{\underline X}

\def\bF{\mathbf F}
\def\bh{\mathbf h}
\def\by{\mathbf y}
\def\bn{\mathbf n}
\def\uy{\underline y}
\def\uY{\underline Y}
\def\uz{\underline z}
\def\bz{\mathbf z}
\def\uv{\underline v}
\def\uth0{{\underline \eta}_0}
\def\dist{\textrm{dist}}
\def\diy{\displaystyle}
\def\ov{\overline}

\def\wh{\widehat}

\def\wtz{\widetilde z}

\def\bI{\mathbf I}
\def\bK{\mathbf K}
\def\bP{\mathbf P}
\def\bV{\mathbf V}
\def\oW{\overline W}
\def\ofW{\overline\fW}
\def\bY{\mathbf Y}
\def\bfB{\mbox{\boldmath${\fB}$}}
\def\bfG{\mbox{\boldmath${\fG}$}}
\def\bfW{\mbox{\boldmath${\fW}$}}

\centerline{\large{\bf A quantum Mermin--Wagner theorem}} 
\centerline{\large{\bf for a generalized Hubbard model}}
\centerline{\large{\bf on a 2D graph}}

\vspace{3mm}
\centerline{\bf Mark Kelbert}

\vspace{1mm}
\centerline{Department of Mathematics, Swansea University, UK}
\vspace{1mm}
\centerline{IME, University 
of S\~ao Paulo, Brazil}
\vspace{1mm}
\centerline{m.kelbert@swansea.ac.uk}
\vspace{3mm}
\centerline{\bf Yurii Suhov}

\vspace{1mm}
\centerline{StatsLab, DPMMS, University of Cambridge, UK}
\vspace{1mm}
\centerline{IME, University 
of S\~ao Paulo, Brazil}
\vspace{1mm}
\centerline{ITP, RAS, Moscow, Russia}
\vspace{1mm}
\centerline{yms@statslab.cam.ac.uk}
\vspace{3mm}

\def\ovr{\overline r}

\def\ttd{\tt d}
\def\ttg{\tt g}
\def\ttG{\tt G}
\def\bttg{\mbox{\boldmath${\ttg}$}}

\def\wt{\widetilde}
\def\wth{\wt h}
\def\wti{\wt i}
\def\wtj{\wt j}
\def\wtx{\wt x}
\def\wty{\wt y}

\def\wtalpha{\widetilde\alpha}
\def\gam{\gamma}
\def\Gam{\Gamma}
\def\oGam{\overline\Gam}
\def\bXi{\mbox{\boldmath${\Xi}$}}
\def\lam{\lambda}
\def\Lam{\Lambda}
\def\bmu{\mbox{\boldmath${\mu}$}}
\def\bnu{\mbox{\boldmath${\nu}$}}
\def\bphi{\mbox{\boldmath${\phi}$}}
\def\vphi{\varphi}
\def\bvphi{\mbox{\boldmath${\vphi}$}}
\def\bpsi{\mbox{\boldmath${\psi}$}}
\def\btau{\mbox{\boldmath${\tau}$}}
\def\om{\omega}
\def\Om{\Omega}
\def\bom{{\mbox{\boldmath$\om$}}}
\def\bOm{\mbox{\boldmath${\Om}$}}

\def\omu{\overline\mu}
\def\obmu{\overline\bmu}
\def\ovphi{\overline\vphi}
\def\obvphi{\overline\bvphi}
\def\obpsi{\overline\bpsi}
\def\utheta{\underline\theta}

\def\oom{\overline\omega}
\def\oOm{\overline\Omega}
\def\sbom{\mbox{\sl${\bom}$}}
\def\obom{\overline\bom}
\def\0bom{{\bom}^0}
\def\0obom{{\obom}^0}
\def\nbom{{\bom}_n}
\def\0nbom{{\bom}_{n,0}}
\def\n*bom{{\bom}^*_{(n)}}

\def\wtbom{\widetilde\bom}
\def\whbom{\widehat\bom}
\def\oom{\overline\om}
\def\wtom{\widetilde\om}

\def\obOm{\overline\bOm}
\def\whbOm{\widehat\bOm}
\def\wtbOm{\widetilde\bOm}

\def\oomega{\overline\omega}
\def\oUpsilon{\overline\Upsilon}
\def\wtomega{\widetilde\omega}
\def\wtheta{\widetilde\theta}

\def\fB{\mathfrak B}
\def\fG{\mathfrak G}
\def\fW{\mathfrak W}

\def\rd{\rm d}
\def\rr{\rm r}
\def\rx{\rm x}
\def\ry{\rm y}
\def\rtr{\rm{tr}}

\def\bcH{\mbox{\boldmath${\cH}$}}

\def\cl{\centerline} 

\def\cA{\mathcal A} 
\def\cB{\mathcal B}
\def\cC{\mathcal C}
\def\cE{\mathcal E} 
\def\cF{\mathcal F} 
\def\cH{\mathcal H} 
\def\cK{\mathcal K} 
\def\cL{\mathcal L} 
\def\cT{\mathcal T}
\def\cV{\mathcal V} 
\def\cW{\mathcal W}
\def\ocW{\overline W}

\def\bbC{\mathbb C}  
\def\bbE{\mathbb E}    
\def\bbP{\mathbb P}
\def\bbQ{\mathbb Q}
\def\bbR{\mathbb R}
\def\bbS{\mathbb S}
\def\bbT{\mathbb T}
\def\bbZ{\mathbb Z}

\def\obbP{\overline\bbP}

\def\oJ{\overline J}
\def\oP{\overline P}

\def\ba{\mathbf a}
\def\bA{\mathbf A}
\def\bB{\mathbf B}
\def\bF{\mathbf F}
\def\bG{\mathbf G}
\def\bH{\mathbf H}
\def\bI{\mathbf I}
\def\bN{\mathbf N}
\def\bR{\mathbf R}
\def\bU{\mathbf U}
\def\bX{\mathbf X}
\def\bx{\mathbf x}
\def\oB{\overline B}
\def\obx{\overline{\mathbf x}}
\def\ox{\overline x}
\def\wtbx{\widetilde\bx}
\def\uk{\underline k}
\def\un{\underline n}
\def\ux{\underline x}
\def\uy{\underline y}
\def\wtux{\widetilde\ux}
\def\uX{\underline X}

\def\bF{\mathbf F}
\def\bh{\mathbf h}
\def\by{\mathbf y}
\def\bn{\mathbf n}
\def\uy{\underline y}
\def\uY{\underline Y}
\def\uz{\underline z}
\def\bz{\mathbf z}
\def\uv{\underline v}
\def\dist{\textrm{dist}}
\def\diy{\displaystyle}
\def\ov{\overline}

\def\wh{\widehat}

\def\wtz{\widetilde z}

\def\bI{\mathbf I}
\def\bK{\mathbf K}
\def\bP{\mathbf P}
\def\bV{\mathbf V}
\def\oW{\overline W}
\def\ofW{\overline\fW}
\def\bY{\mathbf Y}
\def\bfB{\mbox{\boldmath${\fB}$}}
\def\bfG{\mbox{\boldmath${\fG}$}}
\def\bfW{\mbox{\boldmath${\fW}$}}

{\bf Abstract.} This paper is the second in a
series of papers considering symmetry properties of  
bosonic quantum systems over 2D graphs, with continuous
spins, in the spirit of the Mermin--Wagner theorem
\cite{MW}. In the model considered here
the phase space of a single spin is $\cH_1={\rm L}_2(M)$ 
where $M$ is a $d-$dimensional unit torus $M=\bbR^d/\bbZ^d$ 
with a flat metric. The phase space of $k$ spins is 
$\cH_k={\rm L}_2^{\rm{sym}}(M^k)$,
the subspace of ${\rm L}_2(M^k)$ formed by functions symmetric 
under the permutations 
of the arguments. The Fock space 
$\cH=\operatornamewithlimits{\oplus}\limits_{k=0,1,\ldots}\cH_k$
yields the phase space of a system of a varying (but finite) 
number of particles.

We associate a space $\cH\simeq\cH(i)$ with each vertex $i\in\Gam$ of a graph 
$(\Gam ,\cE)$ satisfying a special bi-dimensionality property. (Physically,
vertex $i$ represents a heavy `atom' or `ion' that does not move but attracts
a number of `light' particles.) 
The kinetic energy part of the Hamiltonian includes 
(i) $-\Delta /2$, the minus a
half of the Laplace operator on $M$, responsible for the motion 
of a particle while
`trapped' by a given atom, and (ii) an integral term describing 
possible `jumps'
where a particle may join another atom. The potential part is an operator of 
multiplication by a function (the potential energy of a classical 
configuration)
which is a sum of (a) one-body potentials $U^{(1)}(x)$, $x\in M$, 
describing a field generated by 
a heavy atom, (b) two-body potentials $U^{(2)}(x,y)$, $x,y\in M$, 
showing the interaction between
pairs of particles belonging to the same atom, and (c) two-body 
potentials $V(x,y)$, $x,y\in M$, 
scaled along the graph distance ${\ttd}(i,j)$ between vertices 
$i,j\in\Gam$, which gives the interaction between
particles belonging to different atoms. The system under consideration 
can be considered as a generalized (bosonic) Hubbard model.

We assume that a connected Lie group ${\tt G}$ acts on $M$, 
represented by a Euclidean space or torus of dimension $d'\leq d$, 
preserving the metric and the volume in $M$. Furthermore, we suppose that the
potentials $U^{(1)}$, $U^{(2)}$ and $V$ are ${\tt G}$-invariant. The 
result of the paper is that any (appropriately
defined) Gibbs state generated by the above Hamiltonian is 
${\tt G}$-invariant,  provided
that the thermodynamic variables (the fugacity $z$ and the inverse 
temperature $\beta$) satisfy a certain restriction. This 
extends the Mermin--Wagner-type theorems established in \cite{KS1} 
for a simpler class of quantum 2D models. 
 
As in \cite{KS1}, the definition of a Gibbs state (and its analysis) 
is based on the Feynman--Kac representation for the density matrices.
Following \cite{KS1}, we call such a Gibbs state an FK-DLR state, marking
a connection with DLR-type equations.   

\vskip .5 truecm

\vskip .5 truecm

\vskip 1 truecm
\vskip 2 truecm

\quad{\bf 1. Introduction} 
\vskip .5 truecm

{\bf 1.1. Basic facts on bi-dimensional graphs.} As in \cite{KS1}, we 
suppose that the graph 
$(\Gam ,\cE)$ has been given, with the set of vertices $\Gam$ and 
the set of edges $\cE$. The graph  
has the property that whenever edge $(j',j'')\in\cE$,
the reversed edge $(j'',j')$ belongs in $\cE$ as well. Furthermore, 
graph $(\Gam ,\cE)$ is without multiple edges and  
has a bounded degree, i.e. the number of 
edges $(j,j')$ with 
a fixed initial or terminal vertex is uniformly bounded:
$$\begin{array}{r}
\sup\Big[\max\big(\sharp\,\{j'\in\Gam:\;(j,j')\in\cE\},
\qquad\qquad{}\\
\sharp\,\{j'\in\Gam:\;(j',j)\in\cE\}\big):\;j\in\Gam\Big]
<\infty.\end{array}\eqno{(1.1.1)}$$
The bi-dimensionality 
property is expressed in the bound 
$$0<\sup \left[\frac{1}{n}\,\sharp\,\Sigma (j,n):\;
j\in\Gam ,\,n=1,2,\ldots\right]
<\infty \eqno (1.1.2)$$ 
where $\Sigma (j,n)$ stands for the set of vertices in $\Gam$ at 
the graph distance $n$ from $j\in\Gam$ (a sphere of radius
$n$ around $j$):
$$\Sigma (j,n)=\{j'\in\Gam:\;{\tt d}(j,j')=n\}.\eqno (1.1.3)$$ 
(The graph distance ${\tt d}(j,j')={\tt d}_{\Gam,\cE}(j,j')$ 
between $j,j'\in\Gam$ is determined as the minimal length of a 
path on $(\Gam ,\cE)$ joining $j$ and $j'$.) This implies that
for any $o\in \Gam$ the cardinality $\sharp\,\Lam (o,n)$ of the ball
$$\Lam (o,n)=\{j'\in\Gam:\;{\tt d}(o,j')\leq n\}\eqno (1.1.4)$$ 
grows at most quadratically with $n$.

A justification for putting a quantum system on a graph can be that
graph-like structures become increasingly popular in rigorous Statistical
Mechanics, e.g., in
quantum gravity.  Viz., see \cite{KKP}, \cite{KSY1} and \cite{KSY2}. On the other 
hand, a number of properties of Gibbs ensembles do not depend upon
`regularity' of an underlying  spatial geometry. 
\medskip

{\bf 1.2. A bosonic model in the Fock space.} With each vertex $i\in\Gam$ we 
associate a copy of a compact manifold $M$ which we take in this paper to be 
a unit $d$-dimensional torus $\bbR^d/\bbZ^d$ with a flat metric $\rho$ and the 
volume $v$ . We also associate with $i\in\Gam$
a bosonic Fock--Hilbert space $\bcH (i)\simeq\bcH$. Here 
$\bcH =\operatornamewithlimits{\oplus}\limits_{k=0,1,\ldots}
\cH_k$ where $\cH_k={\rm L}_2^{\rm{sym}}(M^k)$ is the subspace in ${\rm L}_2(M^k)$
formed by functions symmetric under a permutation of the variables.
Given a finite set $\Lam\subset\Gam$, we set $\bcH (\Lam ) =
\operatornamewithlimits{\otimes}\limits_{i\in\Lam}\bcH (i)$. An
element $\bphi\in\bcH (\Lam )$ is a complex function 
$$\bx^*_\Lam\in M^{*\Lam} \mapsto \bphi (\bx^*_{\Lam}).$$
Here $\bx^*_\Lam$ is a collection $\{\bx^*(j),j\in\Lam\}$ 
of finite point sets $\bx^*(j)\subset M$ associated with sites 
$j\in\Lam$. Following \cite{KS1}, we call $\bx^*(j)$ a particle configuration
at site $j$ (which can be empty) and $\bx^*_\Lam$ a particle configuration
in, or over, $\Lam$. The space $M^{*\Lam}$ of particle configurations in $\Lam$
can be represented
as the Cartesian product $(M^*)^{\times\Lam}$ where $M^*$ is the disjoint union
$\operatornamewithlimits{\bigcup}\limits_{k=0,1,\ldots}M^{(k)}$,
$M^{(k)}$ being the collection of (unordered) $k$-point subsets
of $M$. (One can consider $M^{(k)}$ as the factor of the `off-diagonal' 
set $M^k_{\neq}$ in the Cartesian power $M^k$ 
under the equivalence relation induced by the permutation group of order $k$.) 
The norm and the scalar product in $\bcH_\Lam$ are given by
$$\|\bphi \|=\left(\int_{M^{*\Lam}} |\bphi (\bx^*_\Lam)|^2{\rd}\bx^*_\Lam\right)^{1/2}$$
and
$$\langle\bphi_1,\bphi_2\rangle =\int_{M^{*\Lam}} \bphi_1
(\bx^*_\Lam)\ov{\bphi_2 (\bx^*_\Lam)}{\rd}\bx^*_\Lam$$
where measure ${\rd}\bx^*_\Lam$ is the product 
$\operatornamewithlimits{\times}\limits_{j\in\Lam}{\rd}\bx^*(j)$
and ${\rd}\bx^*(j)$ is the Poissonian sum-measure on $M^*$:
$${\rd}\bx^*(j)=\sum\limits_{k=0,1,\ldots}{\mathbf 1}(\sharp\,
\bx^*(j)=k)\frac{1}{k!}\prod\limits_{x\in\bx^*(j)}{\rd}v(x) e^{-v(M)}.$$
Here $v(M)$ is the volume of torus $M$.

As in \cite {KS1}, we assume that an action 
$$({\ttg},x)\in{\ttG}\times M\mapsto {\ttg}\,x\in M,$$
is given, of a group ${\ttG}$ that is a Euclidean space or a
torus of dimension $d'\leq d$. The action is written as
$${\ttg}x=x+\theta A\;{\rm{mod}}\;1.\eqno (1.2.1)$$
Here vector $\theta =(\theta_1,\ldots ,\theta_{d'})$ with components 
$\theta_l\in [0,1)$  and $\theta A$
is the $d$-dimensional vector $\utheta =\left((\theta A)_1,\ldots ,
(\theta A)_{d}\right)$ representing the element ${\ttg}$,
where $A$ is a $(d'\times d)$ matrix of column 
rank $d'$ with rational entries.  
The action of ${\ttG}$ is lifted to unitary operators $\bU_\Lam ({\ttg})$ 
in $\bcH_\Lam$:
$$\bU_\Lam({\ttg})\bphi (\bx^*_\Lam )=\bphi ({\ttg}^{-1}\bx^*_\Lam)
\eqno (1.2.2)$$ 
where ${\ttg}^{-1}\bx^*_\Lam =\{{\ttg}^{-1}\bx^*(j), j\in\Lam\}$ 
and ${\ttg}^{-1}\bx^*(j)=\{{\ttg}^{-1}x,\;x\in\bx^*(j)\}$.

The generally accepted view is that the Hubbard model is Òa highly oversimplified
modelÓ for strongly interacting electrons in a solid. The Hubbard model is a kind
of minimum model which takes into account quantum mechanical motion of electrons in a
solid, and nonlinear repulsive interaction between electrons. There is little doubt that the
model is too simple to describe actual solids faithfully \cite{T}.
In our context the Hubbard Hamiltonian $\bH_\Lam$ of the system in $\Lam$ acts as
follows: 
$$\begin{array}{l}\diy\big(\bH_{\Lam}\bphi\big)(\bx^*_\Lam )=
\bigg[-\frac{1}{2}
\sum\limits_{j\in\Lam}\sum\limits_{x\in\bx^*(j)}\Delta^{(x)}_j
+\sum\limits_{j\in\Lam}\sum\limits_{x\in\bx^*(j)}U^{(1)}(x)\\
\diy\qquad +\frac{1}{2}\sum\limits_{j\in\Lam}\sum\limits_{x,x'\in\bx^*(j)}
{\mathbf 1}(x\neq x')U^{(2)}(x,x')\\
\qquad\diy +\frac{1}{2}\sum\limits_{j,j'\in\Lam}{\mathbf 1}(j\neq j')J({\ttd}(j,j'))
\sum\limits_{x\in\bx^*(j),x'\in\bx^*(j')}V(x,x')\bigg]
\bphi (\bx^*_{\Lam})\\
\qquad\diy+\sum\limits_{j,j'\in\Lam}\lambda_{j,j'}
{\mathbf 1}(\sharp\,\bx^*(j)\geq 1,\;\sharp\,\bx^*(j')<\kappa )\\
\qquad\diy\times\sum\limits_{x\in\bx^*(j)}
\int_Mv({\rd}y)\Big[\bphi\left(\bx^{*(j,x)\to (j',y)}_{\Lam}
\right)
-\bphi (\bx^*_{\Lam})\Big].\end{array}\eqno{(1.2.3)}$$
Here $\Delta^{(x)}_j$ means the Laplacian in variable $x\in
\bx^*(j)$. Next, $\sharp\,\bx^*$ stands for the cardinality of the 
particle configuration $\bx^*$ (i.e., $\sharp\,\bx^*=k$ when $\bx^*\in M^{(k)}$)
and the parameter $\kappa$ is introduced in (1.3.4).\footnote{Symbol $\,\sharp\,$ 
will be used for denoting the cardinality of a general (finite) set;
e.g., $\sharp\,\Lam$ means the number of vertices in $\Lam$.} Further, $\bx^{*(j,x)\to (j',y)}_{\Lam}$ denotes the 
particle configuration with the point $x\in\bx^*(j)$ removed and point 
$y$ added to $\bx^*(j')$.

As in \cite{KS1}, we also consider a Hamiltonian 
$\bH_{\Lam |\obx^*_{{\oGam}\setminus\Lam}}$ in an external field
generated by a configuration $\obx^*_{{\oGam}\setminus\Lam}
=\{\obx^*(j'),\,j'\in{\oGam}\setminus\Lam\}
\in M^{*{\oGam}\setminus\Lam}$ where 
${\oGam}\subseteq\Gam$ is
a (finite or infinite) collection of vertices. More precisely,
we only consider $\obx^*_{{\oGam}\setminus\Lam}$ with $\sharp \obx^*(j')\leq\kappa$
(see Eqn (1.3.4) below) and set 
$$\begin{array}{l}\diy\big(\bH_{\Lam |\obx^*_{{\oGam}\setminus\Lam}}
\bphi\big)(\bx^*_\Lam )=
\bigg[-\frac{1}{2}
\sum\limits_{j\in\Lam}\sum\limits_{x\in\bx^*(j)}\Delta^{(x)}_j
+\sum\limits_{j\in\Lam}\sum\limits_{x\in\bx^*(j)}U^{(1)}(x)\\
\diy\qquad +\frac{1}{2}\sum\limits_{j\in\Lam}\sum\limits_{x,x'\in\bx^*(j)}
{\mathbf 1}(x\neq x')U^{(2)}(x,x')\\
\qquad\diy +\frac{1}{2}\sum\limits_{j,j'\in\Lam}{\mathbf 1}(j\neq j')J({\ttd}(j,j'))
\sum\limits_{x\in\bx^*(j),x'\in\bx^*(j')}V(x,x')\bigg]
\bphi (\bx^*_{\Lam})\\
\diy\qquad +\sum\limits_{j\in\Lam,j'\in{\oGam}\setminus\Lam}J({\ttd}(j,j'))
\sum\limits_{x\in\bx^*(j),\ox'\in\obx^*(j')}V(x,\ox')
\bphi (\bx^*_{\Lam})\\
\qquad\diy+\sum\limits_{j,j'\in\Lam}\lambda_{j,j'}{\mathbf 1}(\sharp\,\bx^*(j)\geq 1,\;
\sharp\,\bx^*(j')<\kappa )\\
\qquad\diy\times\sum\limits_{x\in\bx^*(j)}
\int_Mv({\rd}y)\Big[\bphi\left(\bx^{*(j,x)\to (j',y)}_{\Lam}
\right)-\bphi (\bx^*_{\Lam})\Big].
\end{array}\eqno{(1.2.4)}$$

The novel elements in (1.2.3), (1.2.4) compared with \cite{KS1} 
are the presence of on-site potentials $U^{(1)}$ and $U^{(2)}$ and the 
summand involving transition rates
$\lambda_{j,j'}\geq 0$ for jumps of a particle
from site $j$ to $j'$. 

We will suppose that $\lam_{j,j'}$ vanishes if the graph distance
${\tt d}(j,j')>1$. 
We will also assume uniform boundedness:
$$\sup\,\left[\lam_{j,j'}(x,M),\;j,j'\in\Gam,x\in M\right]<\infty ;
\eqno (1.2.5)$$
in view of (1.1.1) it implies that the total exit rate $\sum\limits_{j':
{\tt d}(j,j')=1}\lam_{j,j'}(x,M)$ from site $j$ is uniformly bounded.
These conditions are not sharp and can be liberalized. 

The model under consideration can be considered as a generalization of
the Hubbard model \cite{H} (in its bosonic version). Its mathematical justification
includes the following. (a) An opportunity to introduce a Fock space formalism 
incorporates a number of new features. For instance, a fermonic version of the model
(not considered here) emerges naturally when the bosonic Fock space $\bcH (i)$ 
is replaced by a fermonic one. Another opening provided by this model is a possibility
to consider random potentials  $U^{(1)}$, $U^{(2)}$ and $V$ which would yield a
sound generalization of the Mott--Anderson model. (b) Introducing  jumps makes
a step towards a treatment of a model of a quantum (Bose-) gas where particles
`live' in a single Fock space. For example, a system of interacting quantum particles 
are originally confined to a `box' in a Euclidean space, with or without `internal' 
degrees of freedom. In the thermodynamical limit the box expands to the whole
Euclidean space.
In a two-dimensional model of a quantum gas one expects a phenomenon of 
invariance under space-translations; one hopes to be able to address this issue
in future publications.  (c) A model with jumps can be analysed by means of
the theory of Markov processes which provides a developed methodology.
 
Physically speaking, the model with jumps covers a situation where `light' quantum 
particles are subject to a `random' force and change their `location'. This class
of models are interesting from the point of view of transport phenomena that they
may display.  (An analogy with the famous Anderson model, in its multi-particle 
version, inevitably comes to mind; 
cf., e.g., \cite{CS}.) Methodologically, such systems 
occupy an `intermediate' place between models where quantum particles are
`fixed' forever to their designated locations (as in \cite{KS1}) and models where
quantum particles move in the same space (a Bose-gas, considered in \cite{SKS1},
\cite{SKS2}). In particular, this work provides a bridge between Refs \cite{KS1} and
\cite{SKS1,SKS2}; reading this paper ahead of   \cite{SKS1,SKS2} might help
an interested reader to get through Refs \cite{SKS1,SKS2} at a much quicker pace.

We would like to note an interesting problem of analysis of the small-mass limit 
(cf. \cite{MVZ}) from the point of Mermin--Wagner phenomena. 

\medskip
 
{\bf 1.3. Assumptions on the potentials.} 
The between-sites potential $V$ is assumed to 
be of class C$^2$. Consequently, $V$ and its  
first and second derivatives satisfy uniform bounds. Viz., $\forall$ $x',x''\in M$
$$
-V(x',x''), \left|\nabla_{{\rx}}V(x',x'')\right|, 
\left|\nabla_{{\rx},{\rx}'}V(x',x'')\right|
\leq {\ov V}.\eqno{(1.3.1)}$$ 
Here ${\rx}$ and ${\rx}'$ run through the pairs of variables $x,x'$.
A similar property is assumed for the on-site potential $U^{(1)}$
(here we need only a $C^1$-smoothness):
$$-U^{(1)}(x), \left|\nabla_{\rx}U^{(1)}(x)\right|\leq {\ov U}^{(1)},
\;\;x\in M.\eqno (1.3.2)$$
Note that for $V$ and $U^{(1)}$ the bounds are imposed on their negative 
parts only.

As to $U^{(2)}$, we suppose that (a)
$$U^{(2)}(x,x')=+\infty\;\hbox{ when $|x-x'|\leq\rho$},\eqno (1.3.3)$$
and (b) $\exists$ a C$^1$-function $(x,x')\mapsto {\wt U}^{(2)}(x,x')\in\bbR$
such that $U^{(2)}(x,x')={\wt U}^{(2)}(x,x')$ whenever $\rho(x,x')>\rho$. 
Here $\rho(x,x')$ stands for the (flat) Riemannian distance between points $x,x'\in M$. 
As a result of (1.3.3), there exists a `hard core' of diameter 
$\rho$, and a given atom cannot `hold' more than 
$$\kappa =v(M)/v(B(\rho))\eqno{(1.3.4)}$$
particles where $v(B(\rho ))$ is the volume of a $d$-dimensional ball
of diameter $\rho$. We will also use the bound
$$-{\wt U}^{(2)}(x,x'), \left|\nabla_{{\rx}}{\wt U}^{(2)}(x,x')\right| 
\leq {\ov U}^{(2)},\;\;x,x'\in M.\eqno (1.3.5)$$
Formally, Eqn (1.3.3) means that the operators  
in (1.2.3) and (1.2.4) are considered for functions $\phi (\bx^*_\Lam )$
vanishing when in the particle configuration $\bx^*_\Lam =\{\bx^* (j),j\in\Lam\}$, 
the  cardinality $\sharp\,\bx^*(j)>\kappa$ for some $j\in\Lam$.  

The function $J:\;r\in (0,\infty )\mapsto
J(r)\geq 0$ is assumed monotonically non-increasing with $r$ and obeying
the relation ${\oJ}(l)\to 0$ as $l\to\infty$ where  
$${\oJ}(l)=\sup\,\left[\sum\limits_{j''\in\Gam}J ({\tt d}(j',j''))
{\mathbf 1}({\tt d}(j',j'')\geq l):\;
j'\in\Gam\right]<\infty .\eqno{(1.3.6)}$$
Additionally, let $J(r)$ be such that
$$ J^*=\sup\left[\sum_{j'\in \Gam} J(d(j,j')) d(j,j')^2: j\in\Gam\right]<\infty.
\eqno (1.3.7)$$
Next, we assume that the functions $U^{(1)}$, $U^{(2)}$ and $V$ 
are ${\ttg}$-invariant: $\forall$ $x,x'\in M$ and ${\ttg}\in{\ttG}$,
$$\begin{array}{c} U^{(1)}(x)=U^{(1)}({\ttg}x),\\
U^{(2)}(x,x')=U^{(2)}({\ttg}x,{\ttg}x'),\\ 
V(x,x')=V({\ttg}x,{\ttg}x').\end{array}\eqno{(1.3.8)}$$
In the following we will need to bound the fugacity (or activity, cf. (1.4.3)) $z$
in terms of the other parameters of the model
$$ze^{\Theta}<1,\;\hbox{ where }\;\Theta
=\kappa\beta \big({\ov U}^{(1)}
+\kappa{\ov U}^{(2)}+\kappa{\oJ}(1){\ov V}\big).\eqno{(1.3.9)}$$
\medskip
{\bf 1.4. The Gibbs state in a finite volume.} Define 
the particle number operator $\bN_\Lam$, with the action
$$\bN_\Lam\bphi (\bx^*_\Lam)=\sharp\,\bx^*_\Lam \bphi 
(\bx^*_\Lam ),\;\;\bx^*_\Lam\in M^{*\Lam}.\eqno (1.4.1)$$
Here, for a given $\bx^*_\Lam =\{\bx^*(j), j\in\Lam\}$, 
$\sharp\,\bx^*_\Lam$ stands for the total number of particles
in configuration $\bx^*_\Lam$:  
$$\sharp\,\bx^*_\Lam =\sum_{j\in\Lam}\sharp\,\bx^*(j).\eqno (1.4.2)$$
The standard canonical variable associated with $\bN_\Lam$
is activity $z\in (0,\infty)$. 

The Hamiltonians (1.2.3), (1.2.4) are self-adjoint  
(on the natural domains) in $\bcH (\Lam )$. 
Moreover, they are positive-definite and have a discrete spectrum.
Furthermore,
$\forall$ $z,\beta >0$,  $\bH_{\Lam}$ and
$\bH_{\Lam |\obx^*_{{\oGam}\setminus\Lam}}$ give rise to 
positive-definite trace-class operators $\bG_\Lam=\bG_{z,\beta,\Lam}$
and $\bG_{\Lam |\obx^*_{{\oGam}\setminus\Lam}}
=\bG_{z,\beta,\Lam |\obx^*_{{\oGam}\setminus\Lam}}$:
$$\bG_\Lam=z^{\bN_\Lam}\exp\,\big[-\beta\bH_{\Lam}\big],\;\;
\bG_{\Lam |\obx^*_{{\oGam}\setminus\Lam}}=z^{\bN_\Lam}\exp\,\left[-\beta\bH_{\Lam |\obx^*_{{\oGam}\setminus\Lam}}\right].\eqno (1.4.3)$$
We would like to stress that the full range of variables $z,\beta >0$ is allowed 
here because of the hard-core condition (1.3.3): it does not allow more than 
$\kappa \sharp\Lam$ particles in $\Lam$ where $\sharp\Lam$ stands for
the number of vertices in $\Lam$. However, while passing to the thermodynamic
limit, we will need to control $z$ and $\beta$. 
\vskip .3 truecm

{\bf Definition 1.1.} 
We will call $\bG_\Lam$ and $\bG_{\Lam |\obx^*_{{\oGam}\setminus\Lam}}$ the Gibbs 
operators in volume $\Lam$, for given values of $z$ and $\beta$ (and -- in the case
of $\bG_{\Lam |\obx^*_{{\oGam}\setminus\Lam}}$ -- with the boundary condition
$\obx^*_{{\oGam}\setminus\Lam}$). 
 
The Gibbs operators in turn give rise to the Gibbs states 
$\bvphi_\Lam =\bvphi_{\beta ,z,\Lam}$ 
and $\bvphi_{\Lam |\obx_{{\oGam}\setminus\Lam}}=
\bvphi_{\beta ,z,\Lam |\obx_{{\oGam}\setminus\Lam}}$, 
at temperature $\beta^{-1}$ and activity $z$ in volume 
$\Lam$. These are linear positive
normalized functionals on the C$^*$-algebra $\bfB_\Lam$ of 
bounded operators in space $\bcH_\Lam$:
$$\bvphi_\Lam (\bA)={\rtr}_{\bcH_{\Lam}}\bR_\Lam\bA,\;
\bvphi_{\Lam |\obx_{{\oGam}\setminus\Lam}} (\bA)
={\rtr}_{\bcH_{\Lam}}\bR_{\Lam |\obx_{{\oGam}\setminus\Lam}}\bA,\;
\;\;\bA\in\fB_\Lam ,\eqno{(1.4.4)}$$
where 
$$\diy
\bR_\Lam =\frac{\bG_\Lam}{\bXi (\Lam )},
\;\hbox{ with }\;
\bXi (\Lam )=\bXi_{z,\beta}(\Lam )
={\rtr}_{{\bcH}_\Lam}\bG_\Lam\eqno (1.4.5)$$
and
$$\begin{array}{l}\diy\bR_{\Lam |\obx^*_{{\oGam}\setminus\Lam}} =
\frac{\bG_{\Lam |\obx_{{\oGam}\setminus\Lam}}}{\bXi (\Lam |\obx^*_{{\oGam}\setminus\Lam})},
\hbox{ with }\;\bXi (\Lam |\obx^*_{{\oGam}\setminus\Lam})
=\bXi_{z,\beta}(\Lam |\obx^*_{{\oGam}\setminus\Lam})\\ 
\qquad\qquad\qquad\qquad\qquad\qquad ={\rtr}_{{\bcH}_\Lam}\Big(z^{\bN_\Lam} 
\exp\,\big[-\beta\bH_{\Lam |\obx^*_{{\oGam}\setminus\Lam}}\big]\Big).
\quad\Diamond\end{array}\eqno (1.4.6)$$
\vskip .3 truecm

Here and below we adopt the following notational agreement: symbol
$\Diamond$ marks the end of a definition, symbol $\lhd$ the end of 
a statement and symbol $\Box$ the end of a proof. 

The hard-core assumption (1.3.3) yields
that the quantities $\bXi (\Lam )$ and $\bXi_{z,\beta}(\Lam |\obx^*_{{\oGam}\setminus\Lam})$
are finite; formally, these facts will be verified by virtue of the Feynman-Kac representation.
\vskip .3 truecm

{\bf Definition 1.2.}
Whenever $\Lambda^0\subset\Lambda$, the
C$^*$-algebra $\bfB_{\Lam^0}$ is identified  
with the C$^*$-sub-algebra in $\bfB_\Lam$ formed by the operators of
the form $\bA_0\otimes\bI_{\Lam\setminus\Lam^0}$. Consequently,
the restriction
$\bvphi_\Lam^{\Lam^0}$ of state $\bvphi_\Lam$ to 
C$^*$-algebra $\bfB_{\Lam^0}$
is given by
$$\bvphi_\Lam^{\Lam^0}(\bA_0)={\rtr}_{\bcH_{\Lam^0}}\big(
\bR_\Lam^{\Lam^0}\bA_0\big),\;\;\bA_0\in\bfB_{\Lam^0}.\eqno{(1.4.7)}$$
where
$$\bR_\Lam^{\Lam^0}=
{\rtr}_{\bcH_{\Lam\setminus\Lam^0}}\bR_\Lam.\eqno{(1.4.8)}$$
Operators $\bR_\Lam^{\Lam^0}$ (we again call them RDMs) are positive-definite and
have ${\rtr}_{\bcH_{\Lam^0}}\bR_\Lam^{\Lam^0}=1$. They also
satisfy the compatibility property: $\forall$ $\Lam^0\subset\Lam^1
\subset\Lam$,
$$\bR_\Lam^{\Lam^0}=
{\rtr}_{\bcH_{\Lam^1\setminus\Lam^0}}\bR_\Lam^{\Lam^1}.
\eqno{(1.4.9)}$$
In a similar fashion one defines functionals
$\bvphi_{\Lam |\obx^*_{{\oGam}\setminus\Lam}}^{\Lam^0}$
and operators 
$\bR_{\Lam |\obx^*_{{\oGam}\setminus\Lam}}^{\Lam^0}$, with 
the same properties. $\quad\Diamond$
\vskip .3 truecm

{\bf 1.5. Limiting Gibbs states.} The concept of a limiting Gibbs state 
is related to notion of a quasilocal $C^*$-algebra. For the class of
systems under consideration, the construction of 
the quasilocal $C^*$-algebra $\bfB_\Gam$
is done along the same lines as in \cite{KS1}: $\bfB_\Gam$
is the norm-completion of the $*$-algebra
$\left({\bfB}^0_\Gam\right)=\diy
{\operatornamewithlimits{\lim\,\rm{ind}}\limits_{n\to\infty}}\;
\bfB_{\Lam_n}$. Any family of positive-definite operators
$\bR^{\Lam^0}$ in spaces $\bcH_{\Lam^0}$ of trace one, where
$\Lam^0$ runs through finite subsets of $\Gam$, with
the compatibility property 
$$\bR^{\Lam^1}=
{\rtr}_{\bcH_{\Lam^0\setminus\Lam^1}}\bR^{\Lam^0},\;\Lam^1\subset\Lam^0,
\eqno{(1.5.1)}$$
determines a state of $\bfB_\Gam$, see \cite{BR}.

Finally, we introduce unitary operators $\bU_{\Lam^0}({\ttg})$,
${\ttg}\in{\ttG}$, in $\bcH_{\Lam^0}$:
$$\bU_\Lam\bphi (\bx^*_{\Lam^0})=\phi ({\ttg}^{-1}\bx^*_{\Lam^0})
\eqno (1.5.2)$$
where
$${\ttg}^{-1}\bx^*_{\Lam^0}=\{{\ttg}^{-1}\bx^*(j), j\in \Lam^0\}
\;\hbox{and}\;{\ttg}^{-1}\bx^*(j)=\{{\ttg}^{-1}x:\,x\in\bx^*(j)\}.
\eqno (1.5.3)$$
\vskip .3 truecm

{\bf Theorem 1.1.} {\sl Assuming the conditions listed above, 
for all $z,\beta\in (0,+\infty )$ satisfying {\rm{(1.3.9)}} 
and a finite
$\Lam^0\subset\Gam$, operators $\bR_\Lam^{\Lam^0}$ 
form a compact sequence in the trace-norm topology in $\bcH_{\Lam^0}$
as $\Lam\nearrow\Gam$. Furthermore, given any family of 
(finite or infinite)
sets ${\oGam} ={\oGam}(\Lam )\subseteq\Gam$ and configurations
$\obx^*_{{\oGam}\setminus\Lam}=\{\obx^*(i),i\in{\oGam}\setminus\Lam\}$ with 
$\sharp\obx^*(i)<\kappa$,
operators $\bR_{\Lam |\obx_{{\oGam}\setminus\Lam}}^{\Lam^0}$ also 
form a compact sequence in the trace-norm topology. Any limit point, $\bR^{\Lam^0}$, for 
$\left\{\bR_{\Lam}^{\Lam^0}\right\}$ or 
$\left\{\bR_{\Lam |\obx^*_{{\oGam}\setminus\Lam}}^{\Lam^0}\right\}$ 
as $\Lam\nearrow\Gam$,
is a positive-definite operator in $\cH(\Lam^0)$ of trace one. Moreover, if 
$\Lam^1\subset\Lam^0$ and $\bR^{\Lam^0}$ and $\bR^{\Lam^1}$ are
the limits for $\bR_\Lam^{\Lam^0}$ and $\bR_\Lam^{\Lam^1}$ or for 
$\bR^{\Lam^0}_{\Lam |\obx^*_{{\oGam}\setminus\Lam}}$ and 
$\bR^{\Lam^1}_{\Lam |\obx^*_{{\oGam}\setminus\Lam}}$ along
the same subsequence $\Lam_s\nearrow\Gam$ then the property {\rm{(1.5.1)}} 
holds true.   

Consequently, 
the Gibbs states $\bvphi_\Lam$ and $\bvphi_{\Lam |\obx^*_{{\oGam}\setminus\Lam}}$ 
form compact sequences as $\Lam\nearrow\Gam$.$\lhd$}
\vskip .3  truecm

{\bf Remark.} In fact, the assertion of Theorem 1.1 holds without assuming the bi-dimensionality 
condition
on graph $(\Gam ,{\mathcal E})$, only under an assumption that the degree of the vertices
in $\Gamma$ is uniformy bounded.
\vskip .3  truecm

{\bf Definition 1.3.} Any limit point $\bvphi$ for states $\bvphi_\Lam$ and $\bvphi_{\Lam |\obx^*_{{\oGam}\setminus\Lam}}$ is called a limiting Gibbs state (for given $z,\beta\in
(0,+\infty )$). $\quad\Diamond$
\vskip .3  truecm

{\bf Theorem 1.2.} {\sl Under the condition {\rm{(1.3.9)}}, any limiting 
point, $\bR^{\Lam^0}$, for 
$\left\{\bR_{\Lam}^{\Lam^0}\right\}$ or
$\left\{\bR_{\Lam |\obx^*_{{\oGam}\setminus\Lam}}^{\Lam^0}\right\}$, 
as $\Lam\nearrow\Gam$,
is a positive- definite operator of trace one commuting
with operators $\bU_{\Lam^0}(\ttg)$: $\forall$ ${\ttg}\in{\ttG}$,
$$\bU_{\Lam^0}({\ttg})^{-1}\bR^{\Lam^0}\bU_{\Lam^0}({\ttg})=
\bR^{\Lam^0}.\eqno (1.5.4)$$
Accordingly, any limiting Gibbs state $\bvphi$ of $\bfB$ determined by 
a family of limiting operators $\bR^{\Lam^0}$ obeying {\rm{(1.5.4)}} 
satisfies the
corresponding invariance property: $\forall$ finite 
$\Lam^0\subset\Gam$, any $\bA\in\bfB_{\Lam^0}$ and ${\ttg}\in{\ttG}$, 
$$\bvphi (\bA)=\vphi (\bU_{\Lam^0}({\ttg})^{-1}
\bA\bU_{\Lam^0}({\ttg})).\lhd\eqno (1.5.5)$$}
\vskip .3 truecm

{\bf Remarks. 1.1.} Condition (1.3.9) does not imply the uniqueness
of an infinite-volume Gibbs state (i.e., absence of phase transitions).  

{\bf 1.2.} Properties (1.5.4) and (1.5.5) are trivially fulfilled for the
limiting points $\bR^{\Lam^0}$ and $\bvphi$ of families $\{\bR^{\Lam^0}_\Lam\}$
and $\{\bvphi_\Lam\}$. However, they require a proof for the limit points
of the families $\{\bR^{\Lam^0}_{\Lam|\obx^*_{\Gam\setminus\Lam}}\}$
and $\{\bvphi_{\Lam|\obx^*_{\Gam\setminus\Lam}}\}$.  
\vskip .5 truecm 

The set of limiting Gibbs states (which is non-empty due to Theorem 1.1), 
is denoted by $\bfG^0$. In the Section 3 we describe a 
class $\bfG\supset\bfG^0$ of states of C$^*$-algebra 
$\bfB$ satisfying the FK-DLR equation, similar to that in \cite{KS1}.
\vskip 2 truecm

{\bf 2. Feynman-Kac representations for the RDM kernels in a finite volume}
\vskip .5 truecm

{\bf 2.1. The representation for the kernels of the Gibbs operators.} 
A starting point for the forthcoming analysis is the Feynman--Kac (FK) 
representation for the kernels $\bK _\Lam (\bx^*_\Lam ,\by^*_\Lam)
={\bK}_{\beta ,z,\Lam}(\bx^*_\Lam,\by^*_\Lam) $ and 
${\bF}_\Lam (\bx^*_\Lam ,\by^*_\Lam)
={\bF}_{\beta ,z,\Lam}(\bx^*_\Lam ,\by^*_\Lam)$ of 
operators $\bG_\Lam$ and $\bR_\Lam$. 
\vskip .5 truecm

{\bf Definition 2.1.} Given $(x,i),(y,j)\in M\times\Gam$,
$\oW^{\,\beta}_{(x,i),(y,j)}$ denotes the space of path, or trajectories,
$\oom=\oom_{(x,i),(y,j)}$ in $M\times\Gam$, of time-length $\beta$, with
the end-points $(x,i)$ and $(y,j)$. Formally, $\oom\in\oW^{\,\beta}_{(x,i),(y,j)}$
is defined as follows:   
$$\begin{array}{c}
\oom :\;\tau\in [0,\beta]\mapsto \oom (\tau )=\left(u\big( \oom ,\tau\big),l
\big(\oom ,\tau\big)\right)\in M\times\Gam ,\\
\oom\;\hbox{ is c\'adl\'ag; }\;\oom (0)=(x,i), \oom (\beta -)=(y,j),\\
\oom\;\hbox{ has finitely many jumps on $[0,\beta ]$;}\\ 
\hbox{if a jump  occurs at time $\tau$ then ${\ttd} \left[l\big(\oom ,
\tau -\big),l\big(\oom ,\tau\big)\right]=1$.}\end{array}\eqno (2.1.1)$$ 
The notation $\oom (\tau )$ and its alternative, $\left(u\big( \oom ,\tau\big),l
\big(\oom ,\tau\big)\right)$, for the position and the index of 
trajectory $\oom$ at time $\tau$ will be employed as equal in rights. We use the term 
the temporal section (or simply the section) of path $\oom$ at time $\tau$. 
$\qquad\Diamond$
\vskip .5 truecm

{\bf Definition 2.2.} Let $\bx^*_\Lam =\{\bx^*(i),i\in\Lam\}\in M^{*\Lam}$ and
$\by^*_\Lam =\{\by^*(j),j\in\Lam\}\in M^{*\Lam}$ be particle configurations
over $\Lam$, with $\sharp\,\bx^*_\Lam =\sharp\,\by^*_\Lam$. A matching (or pairing)
$\gam$ between $\bx^*_\Lam$ and $\by^*_\Lam$ is defined as a collection of pairs
$[(x,i),(y,j)]_\gam$, with $i,j\in\Lam$, $x\in\bx^*(i)$ and $y\in\by^*(j)$, with the 
properties that 
(i) $\forall$ $i\in\Lam$ and $x\in\bx^*(i)$ there exist unique
$j\in\Lam$ and  $y\in\by^*(j)$ such that $(x,i)$ and $(y,j)$ form a pair,
and (ii) $\forall$ $j\in\Lam$ and $y\in\by^*(j)$ there exist unique
$i\in\Lam$ and  $x\in\bx^*(i)$ such that $(x,i)$ and $(y,j)$ form a pair.
(Owing to the condition $\sharp\,\bx^*_\Lam =\sharp\,\by^*_\Lam$, these properties 
are equivalent.) It is convenient to write 
$[(x,i),(y,j)]_\gam =[(x,i),\gam (x,i)]$.

Next, consider the Cartesian product
$$\oW^{\,\beta}_{\bx^*_\Lam ,\by^*_\Lam ,\gam}=
{\operatornamewithlimits{\times}\limits_{i\in\Lam}}\;
{\operatornamewithlimits{\times}\limits_{x\in\bx^*(i)}}
\oW^{\,\beta}_{(x,i),\gam (x,i)}\eqno (2.1.2)$$
and the disjoint union
$$\oW^{\,\beta}_{\bx^*_\Lam ,\by^*_\Lam}=
{\operatornamewithlimits{\bigcup}\limits_{\gam}}\;
\oW^{\beta}_{\bx^*_\Lam ,\by^*_\Lam ,\gam}.\eqno (2.1.3)$$
Accordingly, an element $\obom_\Lam\in \oW^{\,\beta}_{\bx^*_\Lam ,
\by^*_\Lam ,\gam}$ in (2.1.2) represents a collection of paths
$\oom_{x,i}$, $x\in\bx^*(i)$, $i\in\Lam$, of time-length $\beta$, 
starting at $(x,i)$ and ending up 
at $\gam (x,i)$. We say that $\obom_\Lam$ is a path configuration
in (or over) $\Lam$. 
$\quad\Diamond$
\vskip .5 truecm

The presence of matchings in the above construction
is a feature of the bosonic nature of the systems under consideration. 

We will work with standard sigma-algebras (generated by cylinder
sets) in $\oW^{\,\beta}_{(x,i),(y,j)}$, $\oW^{\,\beta}_{\bx^*_\Lam ,\by^*_\Lam,\gam}$ and 
$\oW^{\,\beta}_{\bx^*_\Lam ,\by^*_\Lam}$.
\vskip .5 truecm

{\bf Definition 2.3.} In what follows, $\xi (\tau )$, $
\tau\geq 0$, stands for the Markov process on $M\times\Gam$, with c\'adl\'ag 
trajectories, determind by the generator $\cL$ 
acting on a function $(x,i)\in M\times\Lam\mapsto\psi (x,i)$ by
$$\cL\psi (x,i)=-\frac{1}{2}\Delta\psi (x,i) 
+\sum_{j:{\ttd}(i,j)=1}\lam_{i,j}\int_M v({\rd}y )
\big[\psi (y,j)-\psi (x,i)\big].\Diamond\eqno (2.1.4)$$
In the probabilistic literature, such processes are referred to as L\'evy
processes; cf., e.g., \cite{Sa}.

Pictorially, a trajectory of process $\xi$ moves along $M$ according to 
the Brownian motion with the generator $-\Delta/2$ and changes the index 
$i\in\Gam$ from time to time in 
accordance with jumps occurring in a Poisson process of rate 
$\sum\limits_{j:{\ttd}(i,j)=1}\lam_{i.j}$. In other words, 
while following a Brownian motion rule on $M$,
having index $i\in\Gam$ and being at point $x\in M$, the moving particle 
experiences an urge to jump from $i$ to a neighboring vertex $j$ and 
to a point $y$ at rate $\lam_{i,j}v({\rd}y)$.  After a jump,
the particle continues the Brownian motion on $M$ from $y$ and keeps its 
new index $j$ until the next jump, and so on.

For a given pairs $(x,i),(y,j)\in M\times\Gam$, we denote by 
${\bbP}^{\beta}_{(x,i),(y,j)}$ the non-normalised measure on 
$\oW^{\,\beta}_{(x,i),(y,j)}$ induced by $\xi$. That is, under measure 
${\bbP}^{\beta}_{(x,i),(y,j)}$ the trajectory at time $\tau =0$ starts   
from the point $x$ and has the initial index $i$ while  at time 
$\tau =\beta$ it is at the point $y$ and has the index $j$.
The value ${\wh p}_{(x,i),(y,j)}={\bbP}^{\beta}_{(x,i),(y,j)}\left(\oW^{\,\beta}_{(x,i),(y,j)}\right)$
is given by 
$$\begin{array}{l}{\wh p}_{(x,i),(y,j)}
={\mathbf 1}(i=j)
p_M^\beta(x,y)\exp\,\left[-\beta\sum\limits_{{\wtj}:{\ttd}(i,{\wtj})=1}\lam_{i,{\wtj}}\right]\\
\diy\quad +
\sum\limits_{k\geq 1}
\;\sum\limits_{l_0=i,\,l_1,\ldots ,l_k,l_{k+1}=j}\;\prod_{0\leq s\leq k}{\mathbf 1}
({\ttd}(l_s,l_{s+1})=1)\\
\diy\qquad\times
\;\lam_{\,l_s,l_{s+1}}\int\limits_0^\beta {\rd}\tau_s 
\exp\,\left[-(\tau_{s+1}-\tau_s)
\sum\limits_{{\wtj}:{\ttd}(l_s,{\wtj})=1}\lam_{l_s,{\wtj}}\right]\\
\qquad\times
{\mathbf 1}\Big(0=\tau_0<\tau_1<\ldots <\tau_k<\tau_{k+1}=\beta\Big)\end{array}\eqno (2.1.5)$$
where $p_M^\beta(x,y)$ denotes the transition probability density for the Brownian
motion to pass from $x$ to $y$ on $M$ in time $\beta$:
$$p^\beta_M(x,y)=\frac{1}{(2\pi\beta)^{d/2}}\sum_{\un =(n_1,\ldots ,n_d)\in\bbZ^d} 
\exp\,\big(-|x-y+\un |^2/2\beta\big)\,,\eqno (2.1.6)$$
In view of (1.2.5), the quantity ${\wh p}_{(x,i),(y,j)}$ and its derivatives
are uniformly bounded:
$${\wh p}_{(x,i),(y,j)},\left|\nabla_x{\wh p}_{(x,i),(y,j)}\right|,
\left|\nabla_y{\wh p}_{(x,i),(y,j)}\right|\leq{\wh p}_M,\;x,y\in M,i,j\in\Gam ,
\eqno (2.1.7)$$
where ${\wh p}_M={\wh p}_M(\beta )\in (0,+\infty )$ is a constant.
\vskip 0.5 truecm

We suggest a term 
`non-normalised Brownian bridge with jumps' for the measure
but expect that a better term will be proposed in future.
\vskip 0.5 truecm

{\bf Definition 2.4.} Suppose that 
$\bx^*_\Lam =\{\bx^*(i),i\in\Lam\}\in M^{*\Lam}$ and
$\by^*_\Lam =\{\by^*(j),j\in\Lam\}\in M^{*\Lam}$ are 
particle configurations  
over $\Lam$, with $\sharp\,\bx^*_\Lam =\sharp\,\by^*_\Lam$. Let $\gam$
be a pairing between $\bx^*_\Lam$ and
$\by^*_\Lam$. Then ${\bbP}^{\,*}_{\bx^*_{\Lam},\by^*_{\Lam},\gam}$
denotes the product-measure on $\oW^{\,\beta}_{\bx^*_\Lam ,\by^*_\Lam ,\gam}$:
$${\ov\bbP}^{\,\beta}_{\bx^*_{\Lam},\by^*_{\Lam},\gam}
={\operatornamewithlimits{\times}\limits_{i\in\Lam}}\;
{\operatornamewithlimits{\times}\limits_{x\in\bx^*(i)}}
{\ov\bbP}^{\,\beta}_{(x,i),\gam (x,i)}.\eqno (2.1.8)$$
Furthermore, ${\ov\bbP}^{\,\beta}_{\bx^*_{\Lam},\by^*_{\Lam}}$ stands 
for the sum-measure on $\oW^{\,\beta}_{\bx^*_\Lam ,\by^*_\Lam}$:
$${\ov\bbP}^{\,\beta}_{\bx^*_{\Lam},\by^*_{\Lam}}
=\sum\limits_\gam {\ov\bbP}^{\,\beta}_{\bx^*_{\Lam},\by^*_{\Lam},\gam}.\qquad
\Diamond\eqno (2.1.9)$$
\vskip 0.5 truecm

According to Definition 2.4, under the measure ${\ov\bbP}^{\,\beta}_{\bx^*_{\Lam},\by^*_{\Lam},\gam}$,
the trajectories $\oom_{x,i}\in\oW^{\,\beta}_{(x,i),\gam (x,i)}$ constituting $\obom_\Lam$ 
are independent components. (Here the term independence is used
in the measure-theoretical sense.)

As in \cite{KS1}, we will work with functionals on $\oW^{\,\beta}_{\bx^*_\Lam,\by^*_\Lam ,\gam}$ 
representing integrals along 
trajectories. The first such functional, ${\bh}^{\Lam}(\obom_\Lam)$, is given by
$$\begin{array}{l}\diy{\bh}^{\Lam}(\obom_\Lam)=\sum\limits_{i\in\Lam}\;
\sum\limits_{x\in\bx^*(i)}{\bh}^{x,i}(\oom_{x,i})\\
\diy\quad +\frac{1}{2}\sum\limits_{(i,i')\in\Lam\times\Lam}\;
\sum\limits_{x\in\bx^*(i),x'\in\bx^*(i')}
{\bh}^{(x,i),(x',i')}(\oom_{x,i},\oom_{x',i'}).\end{array}\eqno{(2.1.10)}$$ 
Here, introducing the notation $u_{x,i}(\tau )=u(\oom_{x,i},\tau )$ and $u_{x',i'}(\tau )
=u(\oom_{x',i'},\tau )$
for the positions in $M$ of paths $\oom_{x,i}\in 
\oW^{\,\beta}_{(x,i),\gam (x,i)}$ and $\oom_{x',i'}\in 
\oW^{\,\beta}_{(x',i'),\gam (x',i')}$ at time $\tau$, we define:
$${\bh}^{x,i}(\oom_{x,i})=\diy 
\int_0^{\beta}{\rd}\tau\; U^{(1)}
\Big(u_{i,x}(\tau )\Big).\eqno{(2.1.11)}$$ 
Next, with $l_{x,i}(\tau )$ and $l_{x',i'}(\tau )$
standing for the indices of  $\oom_{x,i}$ and $\oom_{x',i'}$
at time $\tau$, 
$$\begin{array}{l}{\bh}^{(x,i),(x',i')}(\oom_{x,i},\oom_{x',i'})\\
\;\; =\diy 
\int_0^{\beta}{\rd}\tau \bigg[\sum\limits_{j'\in\Gam}U^{(2)}
\Big(u_{i,x}(\tau ),u_{i',x'}(\tau )\Big)
{\mathbf 1}\Big(l_{x,i}(\tau)=j'=l_{x',i'}(\tau)\Big)\\
\diy\qquad\qquad\qquad +\frac{1}{2}\sum\limits_{(j',j'')\in\Gam\times\Gam}J({\ttd}(j',j''))\\
\diy\qquad\qquad\times V\Big(u_{i,x}(\tau )
,u_{i',x'}(\tau )\Big)
{\mathbf 1}\Big(l_{x,i}(\tau))=j'\neq j''=l_{x',i'}(\tau )\Big)\bigg].
\end{array}\eqno{(2.1.12)}$$

Next, consider the functional ${\bh}^{\Lam}(\obom_\Lam 
|\obx^*_{{\oGam}\setminus\Lam})$:
for $\obx^*_{{\oGam}\setminus\Lam}=\{\obx^*(j),\;j\in{\oGam}\setminus\Lam\}$.
As before, we assume that $\sharp\,\obx^*(j)\leq\kappa$. Define:
$${\bh}^{\Lam}(\obom_\Lam |\obx^*_{{\oGam}\setminus\Lam})=
{\bh}^{\Lam}(\obom_\Lam )+
{\bh}^{\Lam}(\obom_\Lam |\,|\obx^*_{{\oGam}\setminus\Lam}).
\eqno (2.1.13)$$
Here ${\bh}^{\Lam}(\obom_\Lam )$ is as in (2.1.10) and 
$$\begin{array}{l}
\diy{\bh}^{\Lam}(\obom_\Lam |\,|\obx^*_{{\oGam}\setminus\Lam})=
\sum\limits_{(i,i')\in\Lam\times ({\oGam}\setminus\Lam )}\;
\sum\limits_{x\in\bx^*(i),\ox'\in\obx^*(i')}
{\bh}^{(x,i),(\ox',i')}(\oom_{x,i},(\ox',i'))\end{array}\eqno (2.1.14)$$
where, in turn, 
$$\begin{array}{l}
\diy{\bh}^{(x,i),(\ox',i')}(\oom_{x,i},(\ox',i'))\\
\diy\quad =\int_0^{\beta}{\rd}\tau \Bigg[U^{(2)}
\Big(u_{i,x}(\tau ),\ox'\Big)
{\mathbf 1}\Big(l_{x,i}(\tau)=i'\Big)\\
\diy\qquad +\sum\limits_{j\in\Gam: j\neq i'}J({\ttd}(j,i'))V\Big(u_{i,x}(\tau )
,\ox'\Big){\mathbf 1}\Big(l_{x,i}(\tau))=j\Big)\Bigg].
\end{array}\eqno{(2.1.15)}$$
The functionals ${\bh}^{\Lam}(\obom_\Lam )$ and 
${\bh}^{\Lam}(\obom_\Lam |\,|\obx^*_{{\oGam}\setminus\Lam})$
are interpreted as energies of path configurations. 
Cf. Eqns (2.1.4), (2.3.8) in \cite{KS1}. 

Finally, we introduce  the indicator functional $\alpha_\Lam (\obom_\Lam)$:
$$\alpha_\Lam (\obom_\Lam )=\begin{cases}1,&\hbox{if index}\;
l_{x,i}(\tau )\in\Lam,\;\forall\;\tau\in [0,\beta ],\\
\;& i\in\Lam\;\hbox{ and }\;x\in\bx^*(i),\\
0,&\hbox{otherwise.}\end{cases}\eqno (2.1.16)$$

It can be derived from known results \cite{RS}, \cite{S1} (for a direct
argument, see \cite{Gi}) that the following assertion holds true:  
\vskip 0.5 truecm

{\bf Lemma 2.1.} {\sl For all $z,\beta >0$ and a finite $\Lam$, the Gibbs 
operators $\bG_\Lam$ and 
$\bG_{\Lam |\obx^*_{{\oGam}\setminus\Lam}}$ act as integral operators 
in $\cH(\Lam)$:
$$
\Big(\bG_\Lam\bphi\Big)(\bx^*_\Lam )=\int_{M^{*\Lam}} 
\prod\limits_{j\in\Lam}\prod\limits_{y\in\by^*(j)}
v({\rd}y) {\bK}_\Lam (\bx^*_\Lam,\by^*_\Lam) 
\bphi (\by^*_\Lam)\eqno (2.1.17)$$
and 
$$\begin{array}{l}
\Big(\bG_{\Lam |\obx^*_{{\oGam}\setminus\Lam}}\bphi\Big)(\bx^*_\Lam )\\
\diy\qquad\quad =\int_{M^{*\Lam}} 
\prod\limits_{j\in\Lam}\prod\limits_{y\in\by^*(j)}
v({\rd}y) {\bK}_\Lam (\bx^*_\Lam,\by^*_\Lam
|\obx^*_{{\oGam}\setminus\Lam}) 
\bphi (\by^*_\Lam).\end{array}\eqno (2.1.18)$$
Moreover, the integral kernels ${\bK}_\Lam (\bx^*_\Lam,\by^*_\Lam)$ 
and ${\bK}_\Lam (\bx^*_\Lam,\by^*_\Lam
|\obx^*_{{\oGam}\setminus\Lam})$ vanish if $\sharp\,\bx^*_{\Lam}\neq\sharp\,\by^*_{\Lam}$.
On the other hand, when $\sharp\,\bx^*_{\Lam}=\sharp\,\by^*_{\Lam}$, the kernels\\
${\bK}_\Lam (\bx^*_\Lam,\by^*_\Lam)$ and ${\bK}_\Lam (\bx^*_\Lam,\by^*_\Lam
|\obx^*_{{\oGam}\setminus\Lam})$
admit the following representations:
$$\begin{array}{l}\bK_\Lam(\bx^*_{\Lam},\by^*_{\Lam}) 
=\diy z^{\sharp\,\bx^*_\Lam }\int_{\oW^{\,*}_{\bx^*_\Lam ,\by^*_\Lam}}
{\ov\bbP}^{\,\beta}_{\bx^*_{\Lam},\by^*_{\Lam}}
({\rd}\obom_\Lam)\alpha_\Lam (\obom_\Lam)\\
\qquad\times\exp\,\big[-{\bh}^{\Lam}(\obom_\Lam)\big],
\end{array}\eqno{(2.1.19)}$$
and
$$\begin{array}{l}
{\bK}_\Lam (\bx^*_\Lam,\by^*_\Lam
|\obx^*_{{\oGam}\setminus\Lam}) 
=\diy z^{\sharp\,\bx^*_\Lam }\int_{\oW^{\,*}_{\bx^*_\Lam ,\by^*_\Lam}}
{\ov\bbP}^{\,\beta}_{\bx^*_{\Lam},\by^*_{\Lam}}
({\rd}\obom_\Lam)\alpha_\Lam (\obom_\Lam)\\
\qquad\times\exp\,\big[-{\bh}^{\Lam}(\obom_\Lam |
\obx^*_{{\oGam}\setminus\Lam})\big].\end{array}\eqno (2.1.20)$$
The ingredients of these representations are determined in  {\rm{(2.1.10)}}--{\rm{(2.1.15)}}.
$\lhd$} 
\vskip .5 truecm

{\bf Remark 2.1.} As before, we stress that, owing to (1.3.3), (1.3.4), a non-zero 
contribution to the integral in the RHS of (2.1.19) 
can only come from a path configuration $\obom_\Lam =\{\oom_{x,i}\}$ 
such that $\forall$ $\tau\in [0,\beta ]$ and $\forall$ $j\in\Gam$, the number 
of paths $\oom_{x,i}$ with index $l_{x,i}(\tau )=j$ is less than 
or equal to $\kappa$. Likewise, the integral in the RHS of (2.1.20) receives
a non-zero contribution only come from configurations 
$\obom_\Lam =\{\oom^*_{x,i}\}$ such that, $\forall$ site $j\in\Gam$, the number 
of paths $\oom_{x,i}$ with index $l_{x,i}(\tau )=j$ plus the cardinality
$\sharp\,\obx^*(j)$ does not exceed $\kappa$. 
\vskip .5 truecm

{\bf 2.2. The representation for the partition function.}
The FK-representations of the partition functions  
$\bXi (\Lam )=\bXi_{\beta ,z}(\Lam )$ in (1.4.5) and 
$\bXi (\Lam |\obx^*_{{\oGam}\setminus\Lam})$ in (1.4.6)
reflect a specific character of the traces ${\rm{tr}}\;\bG_\Lam$ 
and ${\rm{tr}}\;\bG_{\Lam |\obx^*_{{\oGam}\setminus\Lam}}$ in $\cH(\Lam)$.
The source of a complication here is the jump terms in the Hamiltonians $H_\Lam$
and $H_{\Lam |\obx^*_{{\oGam}\setminus\Lam}}$ in Eqns (1.2.3) and (1.2.4), respectively.
In particular, we will have to pass from trajectories of fixed time-length $\beta$ to
loops of a variable time-length. To this end, a given  matching $\gamma$ is decomposed
into a product of cycles, and the trajectories associated with a given
cycle are merged into closed paths (loops) of a time-length multiple of $\beta$.
(A similar construction has been performed in \cite{Gi}.)

To simplify the notation, we omit, wherever possible, the index $\beta$. 
\vskip .5 truecm

{\bf Definition 2.5.} For given $(x,i),(y,j)\in M\times\Gam$, the symbol 
$\oW^{\,*}_{(x,i),(y,j)}$ denotes the disjoint union:
$$\oW^{\,*}_{(x,i),(y,j)}={\operatornamewithlimits{\bigcup}\limits_{k=0,1,\ldots}}\;
\oW^{\,k\beta}_{(x,i),(y,j)}\eqno (2.2.1)$$
In other words, $\oW^{\,*}_{(x,i),(y,j)}$ is the space of paths
$\oOm^*=\oOm^*_{(x,i),(y,j)}$ in $M\times\Gam$, of a variable time-length $k\beta$, where
$k=k(\oOm^*)$ takes values $1,2,\ldots$ and called the length multiplicity,  
with the end-points $(x,i)$ and $(y,j)$. The formal definition 
follows the same line as in Eqn (2.1.1), and we again use
the notation $\oOm^*(\tau )$ and the notation $\left(u\big(\oOm^*,\tau\big),l
\big(\oOm^*,\tau\big)\right)$ for the pair of the position and the index of 
path $\oOm^*$ at time $\tau$. Next, we call the particle
configuration $\{\oOm^*(\tau +\beta m),0\leq m<k(\oOm^*)\}$ the temporal section (or simply the section)
of $\oOm^*$ at time $\tau\in [0,\beta ]$. We also call  $\oOm^*_{(x,i),(y,j)}\in\oW^{\,*}_{(x,i),(y,j)}$
a path (from $(x,i)$ to $(y,j)$).

A particular role will be played by closed paths (loops), with coinciding endpoints
(where $(x,i)=(y,j)$). Accordingly, we denote by $W^{\,*}_{x,i}$ the set
$\oW^{\,*}_{(x,i),(x,i)}$. An element of $W^{\,*}_{x,i}$ is denoted
by $\Om^*_{x,i}$ or, in short, by $\Om^*$ and called a loop at vertex $i$. (The upper
index $^*$ indicates that the length multiplicity is unrestricted.) 
The length multiplicity of 
a loop $\Om^*_{x,i}\in W^{\,*}_{x,i}$ is denoted by $k (\bOm^*_{x,i})$
or $k_{x,i}$. It is instructive to
note that, as topological object, a given loop $\bOm^*$ admits a multiple choice of 
the initial pair $(x,i)$: it can be represented by any pair
$\left(u\big(\Om^*,\tau\big),l
\big(\Om^*,\tau\big)\right)$ at a time $\tau =l\beta$ where $l=1,\ldots ,k(\bOm^*)$.
As above, we use the term the temporal section at time $\tau\in [0,\beta ]$
for the particle configuration $\{\Om^*_{x,i}(\tau +\beta m),0\leq m<k_{x,i}\}$
and employ the alternative notation $(u(\tau +\beta m;\Om^*),l(\tau +\beta m;\Om^*))$ addressing
the position and the index of $\Om^*$ at time $\tau +\beta m\in [0,\beta k(\Om^*)]$.  
$\qquad\Diamond$
\vskip .5 truecm

{\bf Definition 2.6.} Suppose $\bx^*_\Lam =\{\bx^*(i),i\in\Lam\}\in M^{*\Lam}$ 
and\\ $\by^*_\Lam =\{\by^*(j),j\in\Lam\}\in M^{*\Lam}$ are particle configurations
over $\Lam$, with\\ $\sharp\,\bx^*_\Lam =\sharp\,\by^*_\Lam$. Let $\gam$ be a matching 
between $\bx^*_\Lam$ and $\by^*_\Lam$. We consider the Cartesian product
$$\oW^{\,*}_{\bx^*_\Lam ,\by^*_\Lam ,\gam}=
{\operatornamewithlimits{\times}\limits_{i\in\Lam}}\;
{\operatornamewithlimits{\times}\limits_{x\in\bx^*(i)}}
\oW^{\,*}_{(x,i),\gam (x,i)}\eqno (2.2.2)$$
and the disjoint union
$$\oW^{\,*}_{\bx^*_\Lam ,\by^*_\Lam}=
{\operatornamewithlimits{\bigcup}\limits_{\gam}}\;
\oW^{\,*}_{\bx^*_\Lam ,\by^*_\Lam ,\gam}.\eqno (2.2.3)$$
Accordingly, an element $\obOm^*_\Lam\in \oW^{\,*}_{\bx^*_\Lam ,
\by^*_\Lam ,\gam}$ in (2.2.3) represents a collection of paths
$\obOm^*_{x,i}$, $x\in\bx^*(i)$, $i\in\Lam$, of time-length $k\beta$, 
starting at $(x,i)$ and ending up at $(y,j)=\gam (x,i)$. We will 
say that $\obOm^*_\Lam\in\oW^{\,*}_{\bx^*_\Lam ,\by^*_\Lam}$ 
is a path configuration in (or over) $\Lam$. 

Again, loops will have a special role and deserve a
particular notation. Namely, $W^*_{\bx^*_{\Lam}}$ denotes the Cartesian
product:
$$W^*_{\bx^*_\Lam}=\;{\operatornamewithlimits{\times}\limits_{i\in\Lam}}
\;{\operatornamewithlimits{\times}\limits_{x\in\bx^*(i)}}W^*_{x,i}
\eqno (2.2.4)$$
and $W^*_\Lam$ stands for the disjoint union (or equivalently, the Cartesian power):
$$W^*_\Lam ={\operatornamewithlimits{\bigcup}\limits_{\bx^*_\Lam\in M^{*\Lam}}}
W^*_{\bx^*_\Lam}={\operatornamewithlimits{\times}\limits_{i\in\Lam}}W^*_{\{i\}}\;
\hbox{where $W^*_{\{i\}}={\operatornamewithlimits{\bigcup}\limits_{\bx^*\in M^*}}
\left({\operatornamewithlimits{\times}\limits_{x\in\bx^*}}W^*_{x,i}\right)$.}
\eqno (2.2.5)$$

A loop configuration  $\bOm^*=\{\bOm^*(i),i\in\Lam\}
\in W^{\,*}_\Lam$  
over $\Lam$ is a collection  of loop 
configurations at vertices $i\in\Lam$ starting and ending up at particle 
configurations $\bx^*(i)\in M^*$ (note that some of the $\bOm^*(i)$'s 
may be empty). The temporal section (or, in short, the section),
$\bOm^*(\tau )$, of $\bOm^*$ at time $\tau$ is defined as the particle configuration
formed by the points $\Om^*_{x,i}(\tau +\beta m)$ where $i\in\Lam$, $x\in\bx^*(i)$
and $0\leq m<k_{x,i}$.  $\quad\Diamond$
\vskip .5 truecm

As before, we consider the standard sigma-algebras of subsets in the spaces
$\oW^{\,*}_{(x,i),(y,j)}$, $W_{x,i}$, $\oW^{\,*}_{\bx^*_\Lam ,\by^*_\Lam ,\gam}$, 
$\oW^{\,*}_{\bx^*_\Lam ,\by^*_\Lam}$, $W^*_{\bx^*_\Lam}$ and $W^*_\Lam$ introduced 
in Definitions 2.5 and 2.6. In particular, the sigma-algebra of subsets in $W^*_\Lam$
will be denoted by $\bfW_\Lam$; we comment on some of its specific properties
in Section 3.1. (An infinite-volume version $W^*_\Gam$ of
$W^*_\Lam$ is treated in Section 3.2 and after.)
\vskip .5 truecm

{\bf Definition 2.7.} Given points $(x,i),(y,j)\in M\times\Gam$, we denote by 
${\ov\bbP}^{\,*}_{(x,i),(y,j)}$ the sum-measure on $\oW^{\,*}_{(x,i),(y,j)}$:
$${\ov\bbP}^{\,*}_{(x,i),(y,j)}=\sum\limits_{k=0,1,\ldots }
{\ov\bbP}^{\,k\beta}_{(x,i),(y,j)}.\eqno (2.2.6)$$

Further, $\bbP^{\,*}_{x,i}$ denotes the similar measure on $W^*_{x,i}$:
$$\bbP^{\,*}_{x,i}=\sum\limits_{k=0,1,\ldots }
\bbP^{\,k\beta}_{x,i}.\quad\Diamond\eqno (2.2.7)$$
\vskip .5 truecm

{\bf Definition 2.8.} Let $\bx^*_\Lam =\{\bx^*(i),i\in\Lam\}\in M^{*\Lam}$ 
and $\by^*_\Lam =\{\by^*(j), j\in\Lam\}\in M^{*\Lam}$ be  particle 
configurations
over $\Lam$, with $\sharp\,\bx^*_\Lam =\sharp\,\by^*_\Lam$. Let $\gam$ be a matching 
between $\bx^*_\Lam$ and $\by^*_\Lam$, we define the product-measure 
${\ov\bbP}^{\,*}_{\bx^*_\Lam ,\by^*_\Lam ,\gam}$:
$${\ov\bbP}^{\,*}_{\bx^*_\Lam ,\by^*_\Lam ,\gam}=
{\operatornamewithlimits{\times}\limits_{i\in\Lam}}\;
{\operatornamewithlimits{\times}\limits_{x\in\bx^*(i)}}
{\ov\bbP}^{\,*}_{(x,i),\gam (x,i)}\eqno (2.2.8)$$
and the sum-measure
$${\ov\bbP}^{\,*}_{\bx^*_\Lam ,\by^*_\Lam}=\sum\limits_{\gam}\;
{\ov\bbP}^{\,*}_{\bx^*_\Lam ,\by^*_\Lam ,\gam}.\eqno (2.2.9)$$

Next, symbol $\bbP^*_{\bx^*_\Lam}$ stands for the 
product-measure on $W^*_{\bx^*_\Lam}$:
$$\bbP^*_{\bx^*_\Lam}=\;{\operatornamewithlimits{\times}\limits_{i\in\Lam}}
\;{\operatornamewithlimits{\times}\limits_{x\in\bx^*(i)}}\bbP^*_{x,i}.
\eqno (2.2.10)$$

Finally,  ${\rd}\bOm^*_\Lam$ yields the measure on $W^{\,*}_\Lam$:
$${\rd}\bOm^*_\Lam ={\rd}\bx^*_\Lam\times
\bbP^*_{\bx^*_\Lam}({\rd}\bOm^*_\Lam ).  \eqno (2.2.11)$$
Here, for $\bx^*_\Lam =\{\bx^*(i),\;i\in\Lam\}$, we set: ${\rd}\bx^*_\Lam 
=\prod\limits_{i\in\Lam}\prod\limits_{x\in\bx^*(i)}v({\rd}x)$. 
For sites $i$ with $\bx^*(i)=\emptyset$, the corresponding factors
are trivial measures sitting on the empty configurations.
$\quad\Diamond$
\vskip .5 truecm

We again need to introduce energy-type functionals represented by 
integrals along loops. More precisely, we define  
the functionals ${\bh}^\Lam (\bOm^*_\Lam )$ and 
${\bh}^{\Lam}(\bOm^*_\Lam |\obx^*_{{\oGam}\setminus\Lam})$  
which are modifications of the above functionals 
${\bh}^{\Lam}(\obom_\Lam)$ and 
${\bh}^{\Lam}(\obom_\Lam |\obx^*_{{\oGam}\setminus\Lam})$; cf. (2.1.10) 
and (2.1.14).
Say, for a loop configuration $\bOm^*_\Lam =\{\Om^*_{x,i}\}$ 
over $\Lam$ with an initial and final particle configuration $\bx^*_\Lam =\{
\bx^*(i),\;i\in\Lam\}$,
$$\begin{array}{l}\diy{\bh}^\Lam (\bOm^*_\Lam )=
\sum\limits_{i\in\Lam}\;
\sum\limits_{x\in\bx^*(i)}{\bh}^{x,i}(\Om^*_{x,i})\\
\diy\quad +\frac{1}{2}\sum\limits_{(i,i')\in\Lam\times\Lam}\;
\sum\limits_{x\in\bx^*(i),x'\in\bx^*(i')}{\mathbf 1}((x,i)\neq (x',i'))\,
{\bh}^{(x,i),(x',i')}(\Om^*_{x,i},\Om^*_{x',i'}).\end{array}\eqno (2.2.12)$$
To determine the functionals ${\bh}^{x,i}(\Om^*_{x,i})$ and
${\bh}^{(x,i),(x',i')}(\Om^*_{x,i},\Om^*_{x',i'})$, we set, for 
given $m= 0,1,\ldots ,k_{x,i}-1$ and $m'= 0,1,\ldots ,k_{x',i'}-1$:
$$\begin{array}{c}
u_{i,x}(\tau +\beta m)=u(\tau +\beta m;\Om^*_{x,i}),\;
l_{i,x}(\tau +\beta m)=l(\tau +\beta m;\Om^*_{x,i}),\\
u_{i',x'}(\tau +\beta m')=u(\tau +\beta m';\Om^*_{x',i'}),\;\;
l_{i',x'}(\tau +\beta m')=l(\tau +\beta m';\Om^*_{x',i'}).
\end{array}$$
A (slightly) shortened notation $l_{i,x}(\tau +\beta m)$ is used
for the index $l_{x,i}(\tau +\beta m;\Om^*_{x,i})$ and $u_{i,x}(\tau +\beta m)$
for the position $u(\tau +\beta m;\Om^*_{x,i})$ for  
$\Om^*_{x,i}(\tau )\in M\times\Gam$, of the section $\Om^*_{x,i}(\tau )$ of the 
loop $\Om^*_{x,i}$ at time $\tau$, and simiarly with $l_{i',x'}(\tau +\beta m')$ 
and $u_{i',x'}(\tau +\beta m')$. (Note that the pairs $(x,i)$ and $(x',i')$ may
coincide. ) Then
$$\begin{array}{l}{\bh}^{(x,i)}(\Om^*_{x,i})
=\diy\int_0^{\beta}{\rd}\tau\Bigg[\sum\limits_{0\leq m<k_{x,i}}  
U^{(1)}\Big(u_{x,i}(\tau +\beta m)\Big)\\
\quad +\diy\sum\limits_{0\leq m<m'<k_{x,i}}  
\sum\limits_{j\in\Gam}{\mathbf 1}\Big(l_{x,i}(\tau +\beta m)=j=
l_{x,i}(\tau +\beta m')\Big)\\
\qquad\qquad\times U^{(2)}
\Big(u_{x,i}(\tau +\beta m),u_{x,i}(\tau +\beta m')\Big)\Bigg]
\end{array}\eqno{(2.2.13)}$$ 
and
$$\begin{array}{l}{\bh}^{(x,i),(x',i')}(\Om^*_{x,i},\Om^*_{x',i'})\\
\quad =\diy\sum\limits_{0\leq m<k_{x,i}}\sum\limits_{0\leq m'<k_{x',i'}}  
\int_0^{\beta}{\rd}\tau 
\bigg[\sum\limits_{j\in\Gam}\;U^{(2)}
\Big(u_{x,i}(\tau +\beta m),u_{x',i'}(\tau +\beta m')\Big)
\\
\qquad\qquad\times 
{\mathbf 1}\Big(l_{x,i}(\tau +\beta m)=j=
l_{x',i'}(\tau +\beta m')\Big)\\
\diy\qquad +\sum\limits_{(j,j')\in\Gam\times\Gam}J({\ttd}(j,j'))V
\Big(u_{i,x}(\tau +\beta m),u_{i',x'}(\tau +\beta m')\Big)\\
\qquad\qquad\times
{\mathbf 1}\Big(l_{x,i}(\tau +\beta m)=j\neq j'=
l_{x',i'}(\tau +\beta m')\Big)\bigg],\end{array}\eqno{(2.2.14)}$$ 

Next, the functional $B(\bOm^*_\Lam)$ takes into account the bosonic character of the model:
$$B(\bOm^*_\Lam)=\prod_{i\in\Lam}\prod_{x\in\bx^*(i)}\frac{z^{k_{x,i}}}{k_{x,i}}
\,.\eqno (2.2.15)$$
The factor $k_{x,i}^{-1}$ in (2.2.15) reflects the fact that the starting point of a loop
$\Om^*_{x,i}$ may be selected among points $u(\beta m,\;\Om^*_{x,i})$ arbitrarily. 

Next, we  define the functional 
${\bh}^{\Lam}(\bOm^*_\Lam |\obx^*_{{\oGam}\setminus\Lam})$:
for $\obx^*_{{\oGam}\setminus\Lam}=\{\obx^*(j),\;j\in{\oGam}\setminus\Lam\}$,
again assuming that $\sharp\obx^*(j)\leq\kappa$. Set:
$${\bh}^{\Lam}(\bOm_\Lam |\obx^*_{{\oGam}\setminus\Lam})=
{\bh}^{\Lam}(\bOm^*_\Lam )+
{\bh}^{\Lam}(\bOm^*_\Lam |\,|\obx^*_{{\oGam}\setminus\Lam}).\eqno (2.2.16)$$
Here ${\bh}^{\Lam}(\bOm^*_\Lam )$ is as in (2.2.12) and 
$$\begin{array}{l}\diy{\bh}^{\Lam}(\bOm^*_\Lam |\,|\obx^*_{{\oGam}\setminus\Lam})\\
\diy\qquad =
\sum\limits_{(i,i')\in\Lam\times ({\oGam}\setminus\Lam )}\;
\sum\limits_{x\in\bx^*(i),\ox'\in\obx^*(i')}
{\bh}^{(x,i),(\ox',i')}(\Om^*_{x,i},(\ox',i'))\end{array}\eqno (2.2.17)$$
where, in turn, 
$$\begin{array}{l}
\diy{\bh}^{(x,i),(\ox',i')}(\Om^*_{x,i},(\ox',i'))\\
\;\; =\diy\sum\limits_{0\leq m<k_{x,i}} 
\int_0^{\beta}{\rd}\tau \bigg[U^{(2)}\Big(u_{x,i}(\tau +\beta m),x'\Big) 
{\mathbf 1}\Big(l_{x,i}(\tau +\beta m)=i'\Big)\\
\diy\qquad +\sum\limits_{j\in\Gam}J({\ttd}(j,i'))V
\Big(u_{i,x}(\tau +\beta m),x'\Big)
{\mathbf 1}\Big(l_{x,i}(\tau +\beta m)=j\Big)\bigg]. 
\end{array}\eqno{(2.2.18)}$$
As before, the functionals ${\bh}^{\Lam}(\bOm^*_\Lam )$ and 
${\bh}^{\Lam}(\bOm^*_\Lam |\,|\obx^*_{{\oGam}\setminus\Lam})$
have a natural interpretation as energies of loop configurations. 

Finally, as before, the functional $\alpha_\Lam (\bOm^*_\Lam )$
is the indicator that the collection of loops $\bOm^*_\Lam =\{\Om^*_{x,i}\}$ 
does not quit $\Lam$:
$$\alpha_\Lam (\bOm^*_\Lam )=\begin{cases}1,&\hbox{ if }\;\Om^*_{x,i}(\tau )
\in M\times\Lam ,\;\forall\;i\in\Lam,\;x\in\bx^*(i)\\
\;&\hbox{ and }\;0\leq\tau\leq\beta k_{x,i},\\
0,&\hbox{ otherwise.}\end{cases}\eqno (2.2.19)$$

Like above, we invoke known results \cite{RS}, \cite{S1} to establish the 
following statement (again a direct argument can be found in \cite{Gi}):  
\vskip 0.5 truecm

{\bf Lemma 2.2.} {\sl For all finite $\Lam\subset\Gam$ and $z,\beta >0$
satisfying (1.3.9), 
the partition functions $\bXi (\Lam )$ in {\rm{(1.4.5)}} 
and $\bXi (\Lam |\obx^*_{{\oGam}\setminus\Lam})$ in {\rm{(1.4.6)}} admit the 
representations as converging integrals:
$$\bXi (\Lam )=\int_{W^{\,*}_\Lam}{\rd}\bOm^*_\Lam
B(\bOm^*_\Lam )\alpha_\Lam (\bOm^*_\Lam )
\exp\;\left[-{\bh}^\Lam (\bOm^*_\Lam )\right]\eqno (2.2.20)$$
and
$$\begin{array}{l}\bXi (\Lam |\obx^*_{{\oGam}\setminus\Lam})\\
\diy\quad =\int_{W^{\,*}_\Lam}{\rd}\bOm^*_\Lam
B(\bOm^*_\Lam )\alpha_\Lam (\bOm^*_\Lam )
\exp\;\left[-{\bh}^{\Lam}(\bOmË^*_\Lam |\,|\obx^*_{{\oGam}\setminus\Lam})
\right]\end{array}\eqno (2.2.21)$$
with the ingredients introduced in {\rm{(2.2.5)}}--{\rm{(2.2.19)}}.$\quad\lhd$}
\vskip 0.5 truecm

Again, we emphasis that the non-zero contribution to the integral in Eqn (2.2.20) 
can only come from loop configurations $\bOm^*_\Lam =\{\Om^*_{x,i},
i\in\Lam ,x\in\bx^* (i),i\in\Lam\}$ such that $\forall$ vertex 
$j\in\Lam$ and $\tau\in [0,\beta ]$,
the total number of pairs $(u_{x,i}(\tau+m\beta ),l_{x,i}(\tau+m\beta ))$
with $0\leq m<k_{x,i}$, and $l(\tau+m\beta )=j$ does not exceed $\kappa$.
\vskip 0.5 truecm

{\bf Remark 2.2.} The integrals in (2.2.20) and (2.2.21) represent examples of 
partition functions which will be encountered in the forthcoming sections. 
See Eqns (2.3.17), (2.3.18), (2.4.4), (2.4.6), (2.4.8), (2.4.10), (3.1.4) 
and (3.1.6) below.
A general form of such a partition function treated as an integral 
over a set of loop
configurations rather than a trace in a Hilbert space is given in Eqns (2.3.17) 
and (2.3.18). 

\vskip 0.5 truecm

{\bf 2.3. The representation for the RDM kernels.} Let $\Lam^0,\Lam$ be finite sets,
$\Lam^0\subset\Lam\subset\Gam$. The construction 
developed in Section 2.2  also allows us to write a convenient representation
for the integral kernels of the RDMs 
$\bR_\Lam^{\Lam^0}$ (see Eqn (1.4.9)) and 
$\bR^{\Lam^0}_{\Lam|\obx^*_{{\oGam}\setminus\Lam}}$. 
In accordance with Lemma 2.1
and the definition of $\bR_\Lam^{\Lam^0}$ in (1.4.9), the operator
$\bR_\Lam^{\Lam^0}$ acts as an integral operator in $\cH(\Lam^0)$: 
$$\left(\bR_\Lam^{\Lam^0}\bphi\right)(\bx^{*0})
=\int_{M^{*\Lam^0}}\prod\limits_{j\in\Lam^0}\prod\limits_{y\in\by^*(j)}v({\rd}y) 
{\bF}^{\Lam^0}_\Lam (\bx^{*0},\by^{*0})\bphi(\by^{*0})\eqno (2.3.1)$$
where
$$\begin{array}{l}{\bF}^{\Lam^0}_\Lam (\bx^{*0},\by^{*0})\\
\quad :=\diy
\int_{M^{*\Lam\setminus\Lam^0}}\prod\limits_{j\in\Lam\setminus\Lam^0}
\prod\limits_{z\in\bz^*(j)}v({\rd}z) {\bF}_\Lam (\bx^{*0}\vee
\bz^*_{\Lam\setminus\Lam^0},\by^{*0}\vee
\bz^*_{\Lam\setminus\Lam^0})\\ \;\;\\
\diy \quad :=\frac{{\wh\bXi}^{\Lam^0}_\Lam(\bx^{*0},\by^{*0};\Lam\setminus
\Lam^0)}{\bXi (\Lam)}.\end{array}\eqno (2.3.2)$$

We employ here and below the notation $\bx^{*0}$ and $\by^{*0}$
for particle configurations $\bx^*_{\Lam^0}=\{\bx^*(i),\;i\in\Lam^0\}$ 
and $\by^*_{\Lam^0}=\{\by^*(j),\;j\in\Lam^0\}$
over $\Lam^0$. Next, $\bx^{*0}\vee\bz^*_{\Lam\setminus\Lam^0}$;
$\by^{*0}\vee\bz^*_{\Lam\setminus\Lam^0}$ denotes the concatenated 
configurations over $\Lam$. 

Similarly, the RDM $\bR^{\Lam^0}_{\Lam |\obx^*_{{\oGam}\setminus\Lam}}$
is determined by its integral kernel\\ ${\bF}^{\Lam^0}_{\Lam |\obx^*_{{\oGam}
\setminus\Lam}}(\bx^{*0},\by^{*0})$, again admitting the representation
$${\bF}^{\Lam^0}_{\Lam |\obx^*_{{\oGam}
\setminus\Lam}}(\bx^{*0},\by^{*0})
:=\frac{{\wh\bXi}^{\Lam^0}_\Lam(\bx^{*0},\by^{*0};\Lam\setminus
\Lam^0\,|\,\obx^*_{{\oGam}\setminus\Lam})}{\bXi (\Lam)}.\eqno (2.3.3)$$

As in \cite{KS1}, we call ${\bF}^{\Lam^0}_\Lam $ and ${\bF}^{\Lam^0}_{\Lam |\obx^*_{{\oGam}\setminus\Lam}}$ the RDM kernels 
(in short, RDMKs).  The focus of our interest are the numerators 
${\wh\bXi}^{\Lam^0}_\Lam (\bx^{*0},\by^{*0},\Lam\setminus\Lam^0)$ and
${\wh\bXi}^{\Lam^0}_\Lam(\bx^{*0},\by^{*0};\Lam\setminus
\Lam^0|\obx^*_{{\oGam}\setminus\Lam})$ in (2.3.2), (2.3.3).
To introduce the appropriate representation for these quantities, we need
some additional definitions.
\vskip 0.5 truecm

{\bf Definition 2.9.} Repeating (2.2.2)--(2.2.3), symbol  
$\oW^{\,*}_{\bx^{*0},\by^{*0}}$ denotes the disjoint union 
${\operatornamewithlimits{\bigcup}\limits_{\gam^0}}\;
\oW^{\,*}_{\bx^{*0},\by^{*0},\gam^0}$ over matchings 
$\gam^0$  between $\bx^{*0}$ and $\by^{*0}$.  
Accordingly, element $\obOm^{\,*0}=
\obOm^{\,*}_{\bx^{*0},\by^{*0},\gam^0}\in\oW^{\,*}_{\bx^{*0},\by^{*0},,\gam^0}$ 
yields a collection of paths 
$\oOm^*_{(x,i),\gam^0(x,i)}\in\oW^{\,*}_{(x,i),\gamˆ0(x,i)}$ lying in $M\times\Gam$. 
Each path $\oOm^*_{(x,i),\gam^0(x,i)}$ has time-lengths $\beta k_{(x,i),(y,j)}$, begins 
at $(x,i)$ and ends up at $(y,j)=\gam^0(x,i)$ where $x\in\bx^*(i)$, $y\in\by^*(j)$. 
Like above, we will 
use for $\obOm^{*0}$ the term a path configuration over $\Lam^0$. 
Repeating (2.2.8)--(2.2.9), we obtain the measures $\bbP^{\,*}_{\bx^{*0},\by^{*0},\gam^0}$
on $\oW^{\,*}_{\bx^{*0},\by^{*0},\gam^0}$ and $\bbP^{\,*}_{\bx^{*0},\by^{*0}}$
on $\oW^{\,*}_{\bx^{*0},\by^{*0}}$.   
$\quad\Diamond$ 
\vskip 0.5 truecm

The assertion of Lemma 2.3 below again follows directly from 
known results, in conjunction with calculations of the partial trace
${\rm{tr}}_{\cH_{\Lam\setminus\Lam^0}}$ in $\cH_\Lam$. The meaning of 
new ingredients in (2.3.3)--(2.3.6) is explained below.

\vskip 0.5 truecm 

{\bf Lemma 2.3.} {\sl  
The quantity ${\wh\bXi}^{\Lam^0}_\Lam (\bx^{*0},\by^{*0};\Lam\setminus\Lam^0)$
emerging in {\rm{(2.3.2)}} is set to be $0$ when $\sharp\,\bx^{*0}\neq \sharp\,\by^{*0}$. 
On the other hand, for $\sharp\,\bx^{*0}=\sharp\,\by^{*0}$, 
$$\begin{array}{l}{\wh\bXi}^{\Lam^0}_\Lam (\bx^{*0},\by^{*0};\Lam\setminus
\Lam^0)
\diy=\int_{\ocW^{\,*}_{\bx^{*0},\by^{*0}}}
\obbP^{\,*}_{\bx^{*0},\by^{*0}}({\rd}\obOm^{*0})
\oB(\obOm^{*0})\\
\qquad\diy\times\alpha_\Lam (\obOm^{*0}){\mathbf 1}(\obOm^{*0}\in\cF^{\Lam^0})
\exp\left[-{\bh}^{\Lam^0}(\obOm^{*0})\right]{\wh\bXi}^{\Lam^0}_\Lam (\Lam\setminus
\Lam^0|\obOm^{*0})
\end{array}\eqno (2.3.4)$$
where  
$$\begin{array}{l}{\wh\bXi}^{\Lam^0}_\Lam (\Lam\setminus
\Lam^0|\obOm^{*0})=\diy\int_{W^{\,*}_{\Lam\setminus\Lam^0}}{\rd}\bOm^*_{\Lam\setminus\Lam^0}B(\bOm^*_{\Lam\setminus\Lam^0})\\
\qquad\diy\times\alpha_\Lam (\bOm^*_{\Lam\setminus\Lam^0}){\mathbf 1}(\bOm^*_{\Lam\setminus\Lam^0}\in\cF^{\Lam^0})
\exp\;\left[-{\bh}^{\Lam\setminus\Lam^0} (\bOm^*_{\Lam \setminus\Lam^0}|\obOm^{*0})\right].
\end{array}\eqno (2.3.5)$$
Similarly, the quantity
${\wh\bXi}^{\Lam^0}_\Lam(\bx^{*0},\by^{*0};\Lam\setminus
\Lam^0\,|\,\obx^*_{{\oGam}\setminus\Lam})$ from 
{\rm{(2.3.3)}} vanishes when $\sharp\,\bx^{*0}\neq \sharp\,\by^{*0}$. 
For $\sharp\,\bx^{*0}=\sharp\,\by^{*0}$, 
$$\begin{array}{l}
{\wh\bXi}^{\Lam^0}_\Lam(\bx^{*0},\by^{*0};\Lam\setminus
\Lam^0\,|\,\obx^*_{{\oGam}\setminus\Lam})\\
\diy =\int_{\ocW^{\,*}_{\bx^{*0},\by^{*0}}}
\obbP^{\,*}_{\bx^{*0},\by^{*0}}({\rd}\obOm^{*0})
\oB(\obOm^{*0})\\
\qquad\diy\times\alpha_\Lam (\obOm^{*0}){\mathbf 1}(\obOm^{*0}\in\cF^{\Lam^0})
\exp\left[-{\bh}^{\Lam^0}(\obOm^{*0}|\obx^*_{{\oGam}\setminus\Lam})\right]\\
\diy\qquad\qquad\times{\wh\bXi}^{\Lam^0}_\Lam (\Lam\setminus
\Lam^0|\obOm^{*0}\vee\obx^*_{{\oGam}\setminus\Lam})
\end{array}\eqno (2.3.6)$$
where
$$\begin{array}{l}{\wh\bXi}^{\Lam^0}_\Lam (\Lam\setminus
\Lam^0|\obOm^{*0}\vee\obx^*_{{\oGam}\setminus\Lam})=\diy\int_{W^{\,*}_{\Lam\setminus\Lam^0}}{\rd}\bOm^*_{\Lam\setminus\Lam^0}B(\bOm^*_{\Lam\setminus\Lam^0})
\alpha_\Lam (\bOm^*_{\Lam\setminus\Lam^0})\\
\diy\qquad\times{\mathbf 1}(\bOm^*_{\Lam\setminus\Lam^0}\in\cF^{\Lam^0})
\exp\;\left[-{\bh}^{\Lam\setminus\Lam^0} (\bOm^*_{\Lam \setminus\Lam^0}|\obOm^{*0}
\vee\obx^*_{{\oGam}\setminus\Lam})\right].
\end{array}.\eqno (2.3.7)$$
These representations hold $\forall$ $z,\beta >0$ and finite $\Lam^0\subset\Lam\subset\Gam$.
$\quad\lhd$}
\vskip 0.5 truecm

Let us define the functionals  $\oB(\obOm^{*0})$, $\alpha_\Lam (\obOm^{*0})$, 
${\mathbf 1}(\obOm^{*0}\in\cF^{\Lam^0})$, \\ 
${\mathbf 1}(\bOm^*_{\Lam\setminus\Lam^0}\in\cF^{\Lam^0})$
${\bh}^{\Lam^0}(\obOm^{*0})$, ${\bh}^{\Lam\setminus\Lam^0} (\bOm^*_{\Lam \setminus\Lam^0}|\obOm^{*0})$,
${\bh}^{\Lam^0}(\obOm^{*0}|\obx^*_{{\oGam}\setminus\Lam})$ and\\ 
${\bh}^{\Lam\setminus\Lam^0} (\bOm^*_{\Lam \setminus\Lam^0}|\obOm^{*0}
\vee\obx^*_{{\oGam}\setminus\Lam})$
in (2.3.3)--(2.3.7). (The functionals $B(\bOm^*_{\Lam\setminus\Lam^0})$ and $\alpha_\Lam (\bOm^*_{\Lam\setminus\Lam^0})$, are defined as (2.2.15) and (2.2.19), 
respectively, replacing $\Lam$ with $\Lam\setminus\Lam^0$.) 

To this end, let $\obOm^{\,*0}=\obOm^{\,*}_{\bx^{*0},\by^{*0},\gam^0}\in
\oW^{\,*}_{\bx^{*0},\by^{*0},\gam^0}$ be a path configuration  
represented by a collection of paths 
$\oOm^*_{(x,i),\gam^0(x,i)}\in\oW^{\,*}_{(x,i),\gam^{*0}(x,i)}$ ($\oOm^*_{x,i}$ in short),
with end-points $(x,i)$ and $(y,j)=\gam^0(x,i)$, of time-length $\beta k_{(x,i),(y,j)}$.
The functional  
$\oB(\obOm^{*0})$ is given by
$$\oB(\obOm^{*0})=\prod_{i\in\Lam}\prod_{x\in\bx^*(i)}
z^{k_{(x,i),(y,j)}}.\eqno (2.3.8)$$
The functional $\alpha_\Lam (\obOm^{*0})$ is again an indicator:
$$\alpha_\Lam (\obOm^{*0})=\begin{cases}1,&\hbox{ if }\;\Om^*_{(x,i),\gam^0(x,i)}(\tau )
\in M\times\Lam ,\;\forall\;i\in\Lam,\;x\in\bx^*(i)\\
\;&\hbox{ and }\;0\leq\tau\leq\beta k_{(x,i),(y,j)},\\
0,&\hbox{ otherwise.}\end{cases}\eqno (2.3.9)$$

Now let us define the indicator function ${\mathbf 1}(\;\cdot\;\in\cF^{\Lam^0})$ 
in (2.3.4)--(2.3.7). The factor ${\mathbf 1}(\obOm^{*0}\in\cF^{\Lam^0})$ equals 
one iff  every path $\oOm^*_{(x,i),(y,j)}$ from $\obOm^{*0}$, 
of time-length $\beta k_{(x,i),(y,j)}$, starting at $(x,i)\in M\times\Lambda^0$ 
and ending up at $(y,j)=\gam^0(x,i)\in M\times\Lambda^0$ remains in 
$M\times (\Lam\setminus\Lam^0)$ at the intermediate times 
$\beta l$ for $l=1,\ldots ,k_{(x,i),(y,j)}-1$: 
$$\oOm^*_{(x,i),(y,j)}(l\beta )\not\in M\times\Lam^0\;
\forall\;l=1,\ldots ,k_{(x,i),(y,j)}-1$$
(when $k_{(x,i),(y,j)}=1$, this is not a restriction).

Furthermore,  
suppose $\bOm^*_{\Lam\setminus\Lam^0}=\bOm^*_{\bx^*_{\Lam\setminus\Lam^0}}$ is 
a loop configuration over $\Lam\setminus\Lamˆ^0$, with the initial/end configuration
$\bx^*_{\Lam\setminus\Lam^0}=\{\bx^* (i),i\in\Lam\setminus\Lam^0\}$, represented by a 
collection of loops $\Om^*_{x,i},i\in\Lam\setminus\Lam^0,x\in\bx^*(i)$. Then
${\mathbf 1}(\bOm^*_{\Lam\setminus\Lam^0}\in\cF^{\Lam^0})=1$
iff each loop $\Om^*_{x,i}$ of time-length $\beta k_{x,i}$, beginning and finishing at $(x,i)\in M\times (\Lam\setminus\Lam^0)$, does not enter the set $M\times\Lam^0$ at times $\beta l$ for $l=1,\ldots ,k_{x,i}-1$:
$$\Om^*_{x,i}(l\beta )\not\in M\times\Lam^0\;
\forall\;l=1,\ldots ,k_{x,i}-1$$
(again, if $k_{x,i}=1$, this is not a restriction).
\vskip .5 truecm

The functional ${\bh}^{\Lam^0}(\obOm^{*0})$ in Eqn (2.3.4) gives the energy
of the path configuration $\obOm^{*0}$ and is introduced similarly 
to Eqn (2.2.12), {\it mutatis mutandis}.
Next, the functional ${\bh}^{\Lam\setminus\Lam^0}(\bOm^*_{\Lam \setminus\Lam^0} |\obOm^{*0})$ in (2.3.5) represents the energy of the loop configuration $\bOm^*_{\Lam \setminus\Lam^0}$ in the potential field generated by the  path configuration $\obOm^{*0}$:
$$ {\bh}^{\Lam\setminus\Lam^0}( \bOm^*_{\Lam \setminus\Lam^0}|\obOm^{*0})=
{\bh}^{\Lam\setminus\Lam^0}(\bOm^*_{\Lam \setminus\Lam^0})
+\bh (\obOm^{*0}|\,|\bOm^*_{\Lam \setminus\Lam^0}).
\eqno (2.3.10)$$
Here, the summand ${\bh}^{\Lam\setminus\Lam^0}(\bOm^*_{\Lam \setminus\Lam^0})$ yields the energy of
the loop configuration 
$\bOm^*_{\Lam \setminus\Lam^0}$; again cf. (2.2.12). 
Further, the term $\bh (\obOm^{*0}|\,|\bOm^*_{\Lam\setminus\Lam^0})$ yields the energy of interaction between
$\obOm^{*0}$ and $\bOm^*_{\Lam \setminus\Lam^0}$: for a 
path/loop configurations $\obOm^{*0}=\{\oOm^*_{x,i}\}\in\oW^{\,*}_{\bx^{*0},\by^{*0},\gam^0}$ and a $\bOm^*_{\Lam\setminus\Lam^0}=\{\Om^*_{x',i'}\}\in W^*_{\Lam\setminus\Lam^0}$ we set
$$\begin{array}{l}
\bh (\obOm^{*0}|\,|\bOm^*_{\Lam \setminus\Lam^0})\\
\diy\qquad =\sum\limits_{(i,i')\in\Lam^0\times (\Lam\setminus\Lam^0)}\;\sum\limits_{x\in\bx^*(i),x'\in\bx^*(i')}{\bh}^{(x,i),(x',i')}
(\oOm^*_{x,i},\Om^*_{x',i'}).\end{array}\eqno (2.3.11)$$
Here, for a path $\oOm^*_{x,i}=\oOm^*_{(x,i),\gam^0(x,i)}$, of time-length $\beta k_{(x,i),\gam^0(x,i)}$,
and a loop $\Om^*_{x',i'}$, of time-length $\beta k_{x,i}$, 
$$\begin{array}{l}{\bh}^{(x,i),(x',i')}(\oOm^*_{x,i},\Om^*_{x',i'})=\diy\sum\limits_{0\leq m<k_{(x,i),\gam^0(x,i)}}\;\;\;\sum\limits_{0\leq m'<k_{x',i'}}\\
\quad\diy 
\times\int_0^{\beta}{\rd}\tau 
\bigg[\sum\limits_{j\in\Gam}\;U^{(2)}
\Big(u(\tau +\beta m;\oOm^*_{(x,i),\gam^0(x,i)}),u(\tau +\beta m';\Om^*_{x',i'})\Big)
\\
\qquad\times {\mathbf 1}\Big(l_{(x,i),\gam^0(x,i)}(\tau +\beta m)=j
=l_{x',i'}(\tau +\beta m')\Big) \\
\diy\qquad +\sum\limits_{j,j'\in\Gam\times\Gam}J({\ttd}(j,j'))
V\Big(u_{i,x}(\tau +\beta m),u_{i',x'}(\tau +\beta m')\Big)\\
\qquad\times{\mathbf 1}\Big(l_{(x,i),\gam^0(x,i)}(\tau +\beta m)=j\neq j'=
l_{x',i'}(\tau +\beta m')\Big)\bigg].\end{array}
\eqno{(2.3.12)}$$ 
Here, in turn, we employ the shortened notation for the positions
and indices of the sections $\oOm^*_{(x,i),\gam^0(x,i)}(\tau +\beta m)$ and $\Om^*_{x',i'}(\tau +\beta m')$
of $\oOm^*_{(x,i),\gam^0(x,i)}$ and $\Om^*_{x',i'}$ at times $\tau +\beta m$ and $\tau +\beta m'$, 
respectively:    
$$\begin{array}{c}
u_{i,x}(\tau +\beta m)=u(\tau +\beta m;\oOm^*_{(x,i),\gam^0(x,i)}),\\
l_{i,x}(\tau +\beta m)=l(\tau +\beta m;\oOm^*_{(x,i),\gam^0(x,i)}),\\
u_{i',x'}(\tau +\beta m')=u(\tau +\beta m';\Om^*_{x',i'}),\\
l_{i',x'}(\tau +\beta m')=l(\tau +\beta m';\Om^*_{x',i'}).\end{array}$$

Further, the functional ${\bh}^{\Lam^0} (\obOm^{*0}|\obx^*_{{\oGam}\setminus\Lam})$ 
in (2.3.6)   
is determined as in Eqns (2.2.16)--(2.2.18), with  $\oOm^*_{(x,i),\gam^0(x,i)}$
instead of $\Om^*_{x,i}$. Next, for \\
${\bh}^{\Lam\setminus\Lam^0}(\bOm^*_{\Lam\setminus\Lam^0}|
\obOm^{\,*0}\vee\obx^*_{{\oGam}\setminus\Lam})$ in (2.3.7), we set:
$${\bh}^{\Lam\setminus\Lam^0}(\bOm^*_{\Lam\setminus\Lam^0}|
 \obOm^{\,*0}\vee\obx^*_{{\oGam}\setminus\Lam})=
{\bh}^{\Lam\setminus\Lam^0}(\bOm^*_{\Lam\setminus\Lam^0})+
\bh(\bOm^*_{\Lam\setminus\Lam^0}|\,|\obOm^{\,*0}\vee\obx^*_{{\oGam}\setminus\Lam}).
\eqno (2.3.13)$$
Here again, the summand ${\bh}^{\Lam\setminus\Lam^0}(\bOm^*_{\Lam\setminus\Lam^0})$ 
is determined as in (2.2.12). Next, the term ${\bh}^{\Lam\setminus\Lam^0}
(\bOm^*_{\Lam\setminus\Lam^0}|\,
|\obOm^{\,*0}\vee\obx^*_{{\oGam}\setminus\Lam})$
is defined similarly to (2.2.17)--(2.2.18):
$$\begin{array}{l}
{\bh}^{\Lam\setminus\Lam^0}(\bOm^*_{\Lam\setminus\Lam^0}|\,|\obOm^{\,*0}\vee\obx^*_{{\oGam}\setminus\Lam})\\
\diy =\sum\limits_{(i,i')\in (\Lam\setminus\Lam^0)\times\Lam^0}\;
\sum\limits_{x\in\bx^*(i),x'\in\bx^*(i')}
{\bh}^{(x,i),(x',i')}(\Om^*_{x,i},\oOm^*_{x',i'})\\
\diy\quad +\sum\limits_{i\in\Lam\setminus\Lam^0}\;
\sum\limits_{x\in\bx^*(i),\ox'\in\obx^*(i')}
{\bh}^{(x,i),(\ox',i')}(\Om^*_{x,i},(\ox',i'))\end{array}\eqno (2.3.14)$$
with
$$\begin{array}{l}
\diy{\bh}^{(x,i),(x',i')}(\Om^*_{x,i},\oOm^*_{x',i'} )=\sum\limits_{0\leq m<k_{x,i}}\;
\;\sum\limits_{0\leq m'<k_{(x',i'),\gam^0(x',i')}}\\
\diy\quad\times\int_0^{\beta}{\rd}\tau \bigg[\sum\limits_{j\in\Gam}
U^{(2)}\Big(u_{x,i}(\tau +\beta m),u_{x',i'}(\tau +\beta m') \Big) \\
\diy\qquad\times
{\mathbf 1}\Big(l_{x,i}(\tau +\beta m)=j=l_{x',i'}(\tau +\beta m') \Big)\\
\diy\qquad +\sum\limits_{(j,j')\in\Gam\times\Gam}J({\ttd}(j,j'))V
\Big(u_{i,x}(\tau +\beta m),u_{x',i'}(\tau +\beta m')\Big)\\
\diy\qquad\times{\mathbf 1}\Big(l_{x,i}(\tau +\beta m)=
j\neq j'=l_{x',i' }(\tau +\beta m')\Big)\bigg].\end{array}\eqno{(2.3.15)}$$
and
$$\begin{array}{l}
\diy{\bh}^{(x,i),(\ox',i')}(\Om^*_{x,i},(\ox',i'))\\ 
\diy\quad =\sum\limits_{0\leq m<k_{x,i}}\int_0^{\beta}{\rd}\tau \bigg[
U^{(2)}\Big(u_{x,i}(\tau +\beta m),\ox'\Big)
{\mathbf 1}\Big(l_{x,i}(\tau +\beta m)=i'\Big)\\
\diy\qquad +\sum\limits_{j\in\Gam}J({\ttd}(j,i'))V
\Big(u_{i,x}(\tau +\beta m),\ox'\Big){\mathbf 1}\Big(l_{x,i}
(\tau +\beta m)=j)\Big)\bigg]. \end{array}\eqno{(2.3.16)}$$

As before, the functionals ${\bh}^{\Lam}(\bOm^*_\Lam )$ and 
${\bh}^{\Lam}(\bOm^*_\Lam |\,|\obx^*_{{\oGam}\setminus\Lam})$
have a natural interpretation as energies of loop configurations. 

Repeating the above observation, non-zero contributions to the integral in 
(2.3.4) come only from pairs $(\obOm^{*0},\bOm^*_{\Lam\setminus\Lam^0})$ such that
$\forall$ $j\in\Gam$ and $\tau\in [0,\beta ]$,
the total number of pairs  $(u(\tau +\beta m;\oOm^*_{(x,i),\gam^0(x,i)}),
l(\tau +\beta m;\oOm^*_{(x,i),\gam^0(x,i)}))$ with $0\leq m<k_{(x,i),\gam^0(x,i)}$, $i\in\Lam^0$ and $x\in\bx^*(i)$ incident to
the paths of the 
configuration $\obOm^{*0}$ and pairs
$(u(\tau +\beta m';\Om^*_{x',i'}),l(\tau +\beta m';\Om^*_{x',i'}))$ with
$0\leq m'<k_{(x',i')}$, 
$i'\in\Lam\setminus\Lam^0$ and $x'\in\bx^*(i')$ incident to the loops of the 
configuration $\obOm^*_{\Lam\setminus\Lam^0}$ 
does not exceed $\kappa$. Similarly, non-zero contributions to the integral in 
(2.3.6) come only from pairs $(\obOm^{*0},\bOm^*_{\Lam\setminus\Lam^0})$ such that
the above inequality holds when we additionally count points $\ox\in\obx^*(j)$.
\vskip .5 truecm

The integral ${\wh\bXi}^{\Lam^0}_\Lam (\Lam\setminus\Lam^0|\obOm^{*0})$ defined in (2.3.5) 
can be considered as a particular (although important) 
example of a partition function 
in the volume $\Lam\setminus\Lam^0$ with a boundary condition $\obOm^{*0}$. 
Note the presence of the subscript $\Lam$ indicating that the loops 
contributing to ${\wh\bXi}^{\Lam^0}_\Lam (\Lam\setminus\Lam^0|\obOm^{*0})$ can jump 
within volume $\Lam$ only (owing to the indicator functional $\alpha_\Lam$).  
On the other hand, the presence of the indicator functional 
${\mathbf 1}(\bOm^*_{\Lam\setminus\Lam^0}\in\cF^{\Lam^0})$ in the integral
(reflected in the upperscript $\Lam^0$ and the roof sign in the notation ${\wh\bXi}^{\Lam^0}_\Lam (\Lam\setminus\Lam^0|\obOm^{*0})$) 
indicates a particular restriction on the jumps of the loops, forbidding 
them to visit set $\Lam^0$ 
at intermediate times $\beta m$. This is true also for the integral 
${\wh\bXi}^{\Lam^0}_\Lam (\Lam\setminus
\Lam^0|\obOm^{*0}\vee\obx^*_{{\oGam}\setminus\Lam})$ in (2.3.7): it is a 
particular example of a partition function 
in the volume $\Lam\setminus\Lam^0$ with a boundary condition $\obOm^{*0}\vee\obx^*_{{\oGam}\setminus\Lam}$. 

Other useful types of partition functions are $\bXi_{\Gam^0}\left({\wt\Lam}|
\obOm^{*0}\vee\bOm^*_{\Gam^1}\right)$ and\\ $\bXi_{\Gam^0}\left({\wt\Lam}|
\obOm^{*0}\vee\bOm^*_{\Gam^1}\vee\obx^*_{\Gam^2}\right)$ where the sets 
of vertices ${\wt\Lam}$, $\Lam^0$, $\Gam^0$, $\Gam^1$ and $\Gam^2$ satisfy
$${\wt\Lam}\subset\Gam^0\subseteq\Gam,\;\;
\Gam^1,\Gam^2\subset\Gam\setminus{\wt\Lam},\;\;
\Gam^1\cap\Gam^2=\emptyset,\;\;\Lam^0\subset\Gam\setminus ({\wt\Lam}\cup\Gam^1\cup\Gam^2)$$ 
and $\sharp{\wt\Lam},\sharp\Lam^0<+\infty$. Accordingly, $\obOm^{*0}$ is a (finite)   
configuration over $\Lam^0$, $\bOm^*_{\Gam^1}$ a (possibly infinite) loop configuration
over $\Gam^1$ and $\obx^*_{\Gam^2}$ a (possibly infinite) particle configuration over
$\Gam^2$. The partition functions $\bXi_{\Gam^0}\left({\wt\Lam}|
\obOm^{*0}\vee\bOm^*_{\Gam^1}\right)$ and $\bXi_{\Gam^0}\left({\wt\Lam}|
\obOm^{*0}\vee\bOm^*_{\Gam^1}\vee\obx^*_{\Gam^2}\right)$ are given by
$$\begin{array}{l} \bXi_{\Gam^0}\left({\wt\Lam}|\obOm^{*0}\vee
\bOm^*_{\Gam^1}\right)\\
\qquad =\diy\int_{W^{\,*}_{\wt\Lam}}{\rd}\bOm^*_{\wt\Lam}\,
\alpha_{\Gam^0}(\bOm^*_{\wt\Lam})\,B(\bOm^*_{\wt\Lam})\,
\exp\;\left[-{\bh}^{\wt\Lam}(\bOm^*_{\wt\Lam}|\obOm^{*0}\vee\bOm^*_{\Gam^1})\right]
\end{array}\eqno (2.3.17)$$
and 
$$\begin{array}{l} \bXi_{\Gam^0}\left({\wt\Lam}|\obOm^{*0}\vee
\bOm^*_{\Gam^1}\vee\obx^*_{\Gam^2}\right)\\
\qquad=\diy\int_{W^{\,*}_{\wt\Lam}}{\rd}\bOm^*_{\wt\Lam}\,\alpha_{\Gam^0}(\bOm^*_{\wt\Lam})  
\,B(\bOm^*_{\wt\Lam})\,
\exp\;\left[-{\bh}^{\wt\Lam}(\bOm^*_{\wt\Lam}|\obOm^{*0}\vee\bOm^*_{\Gam^1}\vee\obx^*_{\Gam^2})\right]
\end{array}\eqno (2.3.18)$$
with the indicator $\alpha_{\Gam^0}$ as in (2.2.19). These partition functions,
feature loop configurations $\bOm^*_{\wt\lam}$ formed by
loops $\Om^*_{x,i}$, $i\in{\wt\Lam}$, which start and finish in 
${\wt\Lam}$, are confined to $\Gam^0$
and move in a potential field generated by $\obOm^{*0}\vee\bOm^*_{\Gam^1}$ where 
$\obOm^{*0}=\{\oOm^{\,*}_{(x,i),\gam^0(x,i)}\}$ and $\bOm_{\Gam^1}
=\{\Om^*_{x,i},x\in\bx^*(i),i\in\Gam^1\}$
or $\bOm^*_{\Gam^1}\vee\obx^*_{\Gam^2}$ where 
$\obx^*=\{\obx^*(i),i\in\Gam^2\}$. (The latter can be understood as
the concatenation of the loop configuration $\bOm^*_{\Gam^1}$ over $\Gam^1$ and
the loop configuration over $\Gam^2$ formed by the constant trajectories
sitting at points $\ox\in\obx^*(i)$, $i\in\Gam^2$.) 
In (2.3.17) we assume that, $\forall$ $\tau\in [0,\beta ]$ and $j\in\Gam$,  
the number 
$$\begin{array}{c}
\sharp\{(x,i,m):\;i\in{\wt\Lam},l(\tau +m\beta;\Om^*_{x,i})=j,0\leq m<k_{x,i}\}\\
+\sharp\{(x,i):\;i\in\Lam^0,l(\tau +m\beta;\oOm^*_{(x,i),\gam^0(x,i)})=j,0\leq m<k_{(x,i),\gam^0(x,i)}\}\\
+\sharp\{(x,i,m):\;i\in\Gam^1,l(\tau +m\beta;\Om^*_{x,i})=j,0\leq m<k_{x,i}\}
\end{array}$$
does not exceed $\kappa$.
Analogously, in (2.3.18) it is assumed that the same is true for the above 
number plus the cardinality $\sharp\obx^*(j)$.   

Such `modified' partition functions will be used in 
forthcoming sections.
\vskip .5 truecm

{\bf 2.4. The FK-DLR measure $\bmu_\Lam$ in a finite volume.} The Gibbs states $\bvphi_\Lam$ and 
$\bvphi_{\Lam |\obx^*_{{\oGam}\setminus\Lam}}$ give rise to probability measures $\bmu_\Lam$ 
and $\bmu_{\Lam |\obx^*_{{\oGam}\setminus\Lam}}$ on the sigma-algebra  
$\bfW_\Lam$ of subsets of $W^*_\Lam$. The sigma-algebra $\bfW_\Lam$ is constructed 
by following the structure of the space $W^*_\Lam$ (a disjoint union of Cartesian products);
cf. Definition 2.8. The measures $\bmu_\Lam$ and $\bmu_{\Lam |\obx^*_{{\oGam}\setminus\Lam}}$
are determined by their Radon--Nykodym derivative $p_\Lam$ and
$p_{\Lam |\obx^*_{{\oGam}\setminus\Lam}}$
relative to the measure 
${\rd}\bOm^*_\Lam$:
$$\begin{array}{l}
\diy p_\Lam (\bOm^*_\Lam ):=\frac{\bmu_\Lam({\rd}\bOm^*_\Lam)}{{\rd}\bOm^*_\Lam}\\
\diy\qquad =\frac{1}{\Xi (\Lam)}
\alpha_\Lam(\bOm^*_\Lam)B(\bOm^*_\Lam )\exp\,\left[-{\bh}^{\Lam}(\bOm^*_\Lam)\right],
\;\;\bOm^*_\Lam\in W^*_\Lam ,
\end{array}\eqno (2.4.1)$$ 
and 
$$\begin{array}{l}\diy p_{\Lam |\obx^*_{{\oGam}\setminus\Lam}}(\bOm^*_\Lam):=\frac{\bmu_{\Lam |
\obx^*_{{\oGam}\setminus\Lam}}({\rd}\bOm^*_\Lam )}{{\rd}\bOm^*_\Lam}
\frac{1}{\Xi (\Lam |\obx^*_{{\oGam}\setminus\Lam})}\\
\diy\qquad = \alpha_\Lam(\bOm^*_\Lam)B(\bOm^*_\Lam )\exp\,\left[-{\bh}^{\Lam}(\bOm^*_\Lam |
\obx^*_{{\oGam}\setminus\Lam})\right],
\;\;\bOm^*_\Lam\in W^*_\Lam .\end{array}\eqno (2.4.2)$$
Given $\Lam^0\subset\Lam$, the sigma-algebra $\bfW_{\Lam^0}$ is naturally identified
with a sigma-subalgebra of $\bfW_\Lam$; this sigma-subalgebra is again denoted
by $\bfW_{\Lam^0}$. The restrictions of $\bmu_\Lam$ to $\bfW_{\Lam^0}$ and 
$\bmu_{\Lam |\obx^*_{{\oGam}\setminus\Lam}}$ are denoted by 
$\bmu^{\Lam^0}_\Lam$ and $\bmu^{\Lam^0}_{\Lam |\obx^*_{{\oGam}\setminus\Lam}}$; 
these measures are determined by their Radon--Nikodym derivatives $\diy p^{\Lam^0}_\Lam
(\bOm^*_{\Lam^0}):=\frac{\bmu^{\Lam^0}_\Lam({\rd}\bOm^*_{\Lam^0})}{{\rd}\bOm^*_{\Lam^0}}$
and $\diy p^{\Lam^0}_{\Lam |\obx^*_{{\oGam}\setminus\Lam}}(\bOm^*_{\Lam^0})
:=\frac{\bmu^{\Lam^0}_{\Lam |\obx^*_{{\oGam}\setminus\Lam}}({\rd}\bOm^*_{\Lam^0})}{{\rd}\bOm^*_{\Lam^0}}$.

The first key property of the measures $\bmu_\Lam$ and $\bmu_{\Lam |\obx^*_{{\oGam}\setminus\Lam}}$ 
is expressed in the so-called FK-DLR equation. We state it as Lemma 2.4 below; its proof 
repeats a standard argument used in the classical case for establishing the DLR equation 
in a finite volume $\Lam\subset\Gam$.
\vskip .5 truecm

{\bf Lemma 2.4.} {\sl For all $z,\beta >0$ satisfying (1.3.9), and 
$\Lam^0\subset\Lam'\subset\Lam$, the probability 
density $p^{\Lam^0}_\Lam$ admits the form
$$p^{\Lam^0}_\Lam\left(\bOm^*_{\Lam^0}\right)=\int_{ W^{\,*}_{\Lam\setminus\Lam'}}
q^{\Lam^0}_{\Lam\setminus\Lam'}
\left(\bOm^*_{\Lam^0}\big|\bOm^*_{\Lam\setminus\Lam'}\right)
\bmu^{\Lam\setminus\Lam'}_\Lam\left(\rd\bOm^*_{\Lam\setminus\Lam'}\right)\eqno (2.4.3)$$
where
$$\begin{array}{r}\diy
q^{\Lam^0}_{\Lam\setminus\Lam'}\left(\bOm^*_{\Lam^0}\big|\bOm^*_{\Lam\setminus\Lam'}\right) 
=\exp\,\left[-{\bh}^{\Lam^0}\left(\bOm^*_{\Lam^0}\big|\bOm^*_{\Lam\setminus\Lam'}\right)\right]\qquad{}\\
\diy\times\frac{\bXi_\Lam \left( \Lam'\setminus\Lam^0\big|\bOm^*_{\Lam^0}\vee\bOm^*_{\Lam\setminus\Lam'}\right)}{\bXi_\Lam \left( \Lam'\big|\bOm^*_{\Lam\setminus\Lam'}\right)}\end{array}\eqno (2.4.4)$$
and the conditional partition functions $\bXi_\Lam \left( \Lam'\setminus\Lam^0\big|\bOm^*\vee\bOm^*_{\Lam\setminus\Lam'}\right)$ and\\ 
$\bXi_\Lam \left( \Lam'\big|\bOm^*_{\Lam\setminus\Lam'}\right)$ are determined 
as in {\rm{(2.3.17)}}. 

Similarly, for $p^{\Lam^0}_{\Lam |\obx^*_{{\oGam}\setminus\Lam}}$ we have:
$$p^{\Lam^0}_{\Lam |\obx^*_{{\oGam}\setminus\Lam}}\left(\bOm^*_{\Lam^0}\right)
=\int_{W^{\,*}_{\Lam\setminus\Lam'}}q^{\Lam^0}_{\Lam\setminus\Lam'}
\left(\bOm^*_{\Lam^0}\big|\bOm^*_{\Lam\setminus\Lam'}\vee\obx^*_{{\oGam}\setminus\Lam}\right)
\bmu^{\Lam\setminus\Lam'}_{\Lam |\obx^*_{{\oGam}\setminus\Lam}}\left(\rd\bOm^*_{\Lam\setminus\Lam'}\right)\eqno (2.4.5)$$
where
$$\begin{array}{r}\diy
q^{\Lam^0}_{\Lam\setminus\Lam'}\left(\bOm^*_{\Lam^0}\big|\bOm^*_{\Lam\setminus\Lam'}
\vee\obx^*_{{\oGam}\setminus\Lam}\right) 
=\exp\,\left[-{\bh}^{\Lam^0}\left(\bOm^*_{\Lam^0}\big|\bOm^*_{\Lam\setminus\Lam'}
\vee\obx^*_{{\oGam}\setminus\Lam}\right)\right]\\
\diy\times\frac{\bXi_\Lam \left( \Lam'\setminus\Lam^0\big|\bOm^*_{\Lam^0}
\vee\bOm^*_{\Lam\setminus\Lam'}
\vee\obx^*_{{\oGam}\setminus\Lam}\right)}{\bXi_\Lam \left( \Lam'\big|\bOm^*_{\Lam\setminus\Lam'}\vee\obx^*_{{\oGam}\setminus\Lam}\right)}
\qquad{}\end{array}\eqno (2.4.6)$$
and the conditional partition functions $\bXi_\Lam \left( \Lam'\setminus\Lam^0\big|\bOm^*_{\Lam^0}\vee\bOm^*_{\Lam\setminus\Lam'}\vee
\obx^*_{{\oGam}\setminus\Lam}\right)$ and\\ 
$\bXi_\Lam \left( \Lam'\big|\bOm^*_{\Lam\setminus\Lam'}\vee\obx^*_{{\oGam}\setminus\Lam}\right)$ 
are determined as in {\rm{(2.3.17)}}. $\quad\lhd$}
\vskip .5 truecm

As in \cite{KS1}, Eqns (2.4.3) and (2.4.5) mean that the conditional densities\\ $p^{\Lam^0}_\Lam
\left(\bOm^*_{\Lam^0}\big|\bOm^*_{\Lam\setminus\Lam'}\right)$ 
and 
$p^{\Lam^0}_{\Lam |\obx^*_{{\oGam}\setminus\Lam}}\left(\bOm^*_{\Lam^0}\big|\bOm^*_{\Lam\setminus\Lam'}\right)$  
relative to $\sigma$-algebra $\bfW^{\Lam\setminus\Lam'}$ coincide, respectively, with
$q^{\Lam^0}_{\Lam\setminus\Lam'}\left(\bOm^{*0}\big|\bOm^*_{\Lam\setminus\Lam'}\right)$,
and
$q^{\Lam^0}_{\Lam\setminus\Lam'}\left(\bOm^{*0}\big|\bOm^*_{\Lam\setminus\Lam'}
\vee\obx^*_{{\oGam}\setminus\Lam}\right)$, 
for $\bmu_\Lam^{\Lam\setminus\Lam'}$- and $\bmu_{\Lam |\obx^*_{{\oGam}\setminus\Lam}}^{\Lam\setminus\Lam'}$-a.a. 
$\bOm^*_{\Lam\setminus\Lam'}\in W^{*}_{\Lam\setminus\Lam'}$ and a.a. $\bOm^*_{\Lam^0}\in W^{\,*}_{\Lam^0}$.

As in \cite{KS1}, we call the expressions 
$q^{\Lam^0}_{\Lam\setminus\Lam'}\left(\bOm^*_{\Lam^0}\big|\bOm^*_{\Lam\setminus\Lam'}\right)$
and \\
$q^{\Lam^0}_{\Lam\setminus\Lam'}\left(\bOm^*_{\Lam^0}\big|\bOm^*_{\Lam\setminus\Lam'}
\vee\obx^*_{{\oGam}\setminus\Lam}\right)$, as well as the expressions  
${\wh q}^{\Lam^0}_{\Lam\setminus\Lam'}\left(\obOm^{*0}\big|
\bOm^*_{\Lam\setminus\Lam'}\right)$ and ${\wh q}^{\Lam^0}_{\Lam\setminus\Lam'}\left(\obOm^{*0}\big|
\bOm^*_{\Lam\setminus\Lam'}\vee\obx^*_{{\oGam}\setminus\Lam}\right)$ appearing below, 
the (conditional) RDM functionals (in brief, the RDMFs). The same name will be used for 
the quantity $q^{\Lam^0}_{\Lam\setminus\Lam'}\left(\bOm^*_{\Lam^0}\big|\bOm^*_{\Lam\setminus\Lam'}\right)$
from Eqns (3.1.3)--(3.1.4) and the quantity\\ 
${\wh q}^{\Lam^0}_{\Gam\setminus\Lam}\left(\obOm^{*0}\big|
\bOm^*_{\Gam\setminus\Lam}\right)$ from Eqns (3.1.5)--(3.1.6).   
 
The second property is that the RDMKs 
${\bF}^{\Lam^0}_\Lam (\bx^{*0},\by^{*0})$ and\\
${\bF}^{\Lam^0}_{\Lam |\obx^*_{{\oGam}\setminus\Lam}}(\bx^{*0},\by^{*0})$ are
related to the measures $\bmu_\Lam$ and 
$\bmu_{\Lam |\obx^*_{{\oGam}\setminus\Lam'}}$. Again, 
the proof of this fact is done by inspection.
\vskip .5 truecm

{\bf Lemma 2.5.} {\sl The RDMK ${\bF}^{\Lam^0}_\Lam(\bx^{*0},\by^{*0})$ 
is expressed as follows: $\forall$ $\Lam^0\subset\Lam'\subset\Lam$,
$$\begin{array}{l}\diy
{\bF}^{\Lam^0}_\Lam (\bx^{*0},\by^{*0})=
\int_{\ocW^{\,*}_{\bx^0,\by^0}}\obbP^{\,*}_{\bx^0,\by^0}
(\rd \obOm^{*0})\alpha_\Lam(\obOm^{*0})\oB (\obOm^{*0})
{\mathbf 1}\left(\obOm^{*0}\in\cF^{\Lam^0}\right)\\
\diy\qquad\times\int_{W^{\,*}_{\Lam\setminus\Lam'}}
\bmu^{\Lam\setminus\Lam'}\left(\rd\bOm^*_{\Lam\setminus\Lam'}\right)
{\mathbf 1}(\bOm^*_{\Lam\setminus\Lam'}\in
\cF^{\Lam^0}) 
{\wh q}^{\Lam^0}_{\Lam\setminus\Lam'}\left(\obOm^{*0}\big|
\bOm^*_{\Lam\setminus\Lam'}\right) \end{array}\eqno (2.4.7)$$
where
$$\begin{array}{r}\diy
{\wh q}^{\Lam^0}_{\Lam\setminus\Lam'}\left(\obOm^{*0}\big|
\bOm^*_{\Lam\setminus\Lam'}\right) 
=\exp\,\left[-{\bh}^{\Lam^0}\left(\obOm^{*0}\big|
\bOm^*_{\Lam\setminus\Lam'}\right)\right]\qquad{}\\
\diy\times\frac{{\wh\bXi}^{\Lam^0}_\Lam \left( \Lam'\setminus\Lam^0\big|\obOm^{*0}\vee\bOm^*_{\Lam\setminus\Lam'}\right)}{\bXi_\Lam \left( \Lam'\big|\bOm^*_{\Lam\setminus\Lam'}\right)}.\end{array}\eqno (2.4.8)$$

Similarly, 
$$\begin{array}{l}\diy
{\bF}^{\Lam^0}_{\Lam |\obx^*_{{\oGam}\setminus\Lam}} (\bx^{*0},\by^{*0})=
\int_{\ocW^{\,*}_{\bx^0,\by^0}}\obbP^{\,*}_{\bx^0,\by^0}
(\rd \obOm^{*0})\alpha_\Lam(\obOm^{*0})\oB (\obOm^{*0})
{\mathbf 1}\left(\obOm^{*0}\in\cF^{\Lam^0}\right)\\
\diy\qquad\times\int_{W^{\,*}_{\Lam\setminus\Lam'}}
\bmu^{\Lam\setminus\Lam'}_{\Lam |\obx^*_{{\oGam}\setminus\Lam}}
\left(\rd\bOm^*_{\Lam\setminus\Lam'}\right)
{\mathbf 1}(\bOm^*_{\Lam\setminus\Lam'}\in
\cF^{\Lam^0}) 
{\wh q}^{\Lam^0}_{\Lam\setminus\Lam'}\left(\obOm^{*0}\big|
\bOm^*_{\Lam\setminus\Lam'}\vee\obx^*_{{\oGam}\setminus\Lam}\right) 
\end{array}\eqno (2.4.9)$$
where
$$\begin{array}{r}\diy
{\wh q}^{\Lam^0}_{\Lam\setminus\Lam'}\left(\obOm^{*0}\big|
\bOm^*_{\Lam\setminus\Lam'}\obx^*_{{\oGam}\setminus\Lam}\right) 
=\exp\,\left[-{\bh}^{\Lam^0}\left(\obOm^{*0}\big|\bOm^*_{\Lam\setminus\Lam'}
\vee\obx^*_{{\oGam}\setminus\Lam}\right)\right]\qquad{}\\
\diy\times\frac{{\wh\bXi}^{\Lam^0}_\Lam \left( \Lam'\setminus\Lam^0\big|\obOm^{*0}\vee\bOm^*_{\Lam\setminus\Lam'}
\vee\obx^*_{{\oGam}\setminus\Lam}\right)}{\bXi_\Lam \left( \Lam'\big|\bOm^*_{\Lam\setminus\Lam'}\vee\obx^*_{{\oGam}\setminus\Lam}
\right)}.\end{array}\eqno (2.4.10)$$
Here the partition functions ${\wh\bXi}^{\Lam^0}_\Lam \left( \Lam'\setminus\Lam^0\big|\obOm^{*0}\vee\bOm^*_{\Lam\setminus\Lam'}\right)$ 
and\\ ${\wh\bXi}^{\Lam^0}_\Lam \left( \Lam'\setminus\Lam^0\big|
\obOm^{*0}\vee\bOm^*_{\Lam\setminus\Lam'}
\vee\obx^*_{{\oGam}\setminus\Lam}\right)$ are determined in (2.3.5) and (2.3.7).$\quad\lhd$}
\vskip .5 truecm

{\bf Remark 2.3.} Summarizing above observations, the measures $\bmu_\Lam$ and
$\mu_{\Lam |\obx^*_{\Gam'\setminus\Lam}}$ are concentrated on the subset in $W^*_\Lam$
formed by loop configurations $\bOm^*_\Lam$ such that $\forall$ $\tau\in [0,\beta]$,
the section $\bOm^*_\Lam (\tau)$ has $\leq\kappa$ particles at each vertex $i\in\Lam$. 
\vskip 2 truecm

{\bf 3. The class of Gibbs states $\bfG$ for the Fock space model}
\vskip .5 truecm

{\bf 3.1. Definition of class $\bfG$.} In this section we apply the idea from \cite{KS1} 
to define the class of states $\bfG =\bfG_{z,\beta}$ for the model introduced in Section 2
and state a number of results. These results will hold under condition (1.5.3)
which is assumed from now on. As 
in \cite{KS1}, the definition
of a state $\bvphi\in\bfG$ is based on the notion of an FK-DLR probability measure $\bmu$ 
on the space $W^*_\Gam$; the class of these measures will be also denoted by $\bfG$.
\vskip .5 truecm

{\bf Definition 3.1.} Space $W^*_\Gam$ is the (infinite) Cartesian product 
${\operatornamewithlimits{\times}\limits_{i\in\Gam}}W^*_{\{i\}}$ (cf. Eqn (2.2.5));
its elements are loop configurations $\bOm^*_\Gam =\{\bOm^*(i),\;i\in\Gam\}$ over 
$\Gam$. A component $\bOm^*(i)$ is a finite loop configuration
(possibly, empty), with an initial/final particle configuration 
$\bx^*(i)\subset M$. Formally, $\bOm^*(i)$ is a finite collection of loops $\Om^*_{x,i}$, of time-length $\beta k_{x,i}$ where 
$k_{x,i}=1,2,\ldots$, starting and finishing at a point 
$(x,i)\in M\times\Gam$. For reader's convenience, we repeat Eqn (2.1.1) 
for the case under consideration:  
$$\begin{array}{c}
\Om^*_{x,i}:\;\tau\in [0,\beta k_{x,i}]\mapsto \left({\wtx}\big( \Om^*_{x,i},\tau\big),{\wti}\big(\Om^*_{x,i},\tau\big)\right)\in M\times\Gam ,\\
\Om^*_{x,i}\;\hbox{ is c\'adl\'ag; }\;\Om^*_{x,i}(0)=\Om^*_{x,i}(\beta k_{x,i}-)=(x,i),\\
\Om^*_{x,i}\;\hbox{ has finitely many jumps on $[0,\beta k_{x,i}]$;}\\ 
\hbox{if a jump  occurs at time $\tau$ then ${\ttd} \left[{\wti}\big(\Om^*_{x,i},\tau -\big),{\wti}\big(\Om^*_{x,i},\tau\big)\right]=1$.}
\end{array}\eqno (3.1.1)$$ 

By $\bfW=\bfW_\Gam$ we denote the $\sigma$-algebra in $W^{*}_\Gam$ generated by cylindrical events. Given a subset ${\oGam}\subset\Gam$ (finite or infinite), we denote by $\bfW^{{\oGam}}=\bfW_{\Gam}^{{\oGam}}$ the 
$\sigma$-subalgebra of $\bfW$ generated by cylindrical events localized in ${\oGam}$. Given a probability  measure $\bmu=\bmu_\Gam$ on $( W^*_\Gam,\bfW_\Gam )$, we denote by $\bmu^{{\oGam}}=\bmu_{\Gam}^{{\oGam}}$ the restriction of $\bmu$
on $\bfW^{{\oGam}}$. $\quad\Diamond$ 
\vskip .5 truecm

{\bf Definition 3.2.}  The class $\bfG$ under consideration 
is formed by  measures $\bmu$ which satisfy the following equation:
$\forall$ finite $\Lam\subset\Gam$ and $\Lam^0\subseteq\Lam$, the probability density 
$$p^{\Lam^0}\left(\bOm^{*0}\right)
=p^{\Lam^0}_{\bmu} \left(\bOm^{*0}\right):=\diy\frac{\bmu^{\Lam^0}_\Gam \left({\rd}\bOm^{*0}\right)}{\bnu\left(\rd\bOm^{*0}\right)},\;\;\bOm^{*0}
\in W^*_{\Lam^0},\eqno (3.1.2)
$$ 
is of the form
$$
p^{\Lam^0}\left(\bOm^*_{\Lam^0}\right)=\int_{ W^{\,*}_{\Gam\setminus\Lam}}
q^{\Lam^0}_{\Gam\setminus\Lam}
\left(\bOm^{*0}\big|\bOm^*_{\Gam\setminus\Lam}\right)
\bmu^{\Gam\setminus\Lam}\left(\rd\bOm^*_{\Gam\setminus\Lam}\right)\eqno (3.1.3)$$
where
$$\begin{array}{r}\diy
q^{\Lam^0}_{\Gam\setminus\Lam}\left(\bOm^{*0}\big|\bOm^*_{\Gam\setminus\Lam}\right) 
=\exp\,\left[-{\bh}^{\Lam^0}\left(\bOm^{*0}\big|\bOm^*_{\Gam\setminus\Lam}\right)\right]\qquad{}\\
\diy\times\frac{\bXi_\Gam \left( \Lam\setminus\Lam^0\big|\bOm^{*0}\vee\bOm^*_{\Gam\setminus\Lam}\right)}{\bXi_\Gam \left( \Lam\big|\bOm^*_{\Gam\setminus\Lam}\right)}\end{array}\eqno (3.1.4)$$
and the conditional partition functions $\bXi_\Gam \left( \Lam\setminus\Lam^0\big|\bOm^{*0}\vee\bOm^*_{\Gam\setminus\Lam}\right)$ and\\ 
$\bXi_\Gam \left( \Lam\big|\bOm^*_{\Gam\setminus\Lam}\right)$ are determined as in (2.3.17).
$\quad\Diamond$ 
\vskip .5 truecm

As in \cite{KS1}, Eqn (3.1.3) means that the conditional density\\ $p^{\Lam^0|\Gam\setminus\Lam}\left(\bOm^{*0}\big|\bOm^*_{\Gam\setminus\Lam}\right)$, 
relative to $\sigma$-algebra $\bfW^{\Gam\setminus\Lam}$, coincides with\\
$q^{\Lam^0}_{\Gam\setminus\Lam}\left(\bOm^{*0}\big|\bOm^*_{\Gam\setminus\Lam}\right)$, 
for $\bmu_\Gam^{\Gam\setminus\Lam}$-a.a.
$\bOm^*_{\Gam\setminus\Lam}\in W^{*\beta}_{\Gam\setminus\Lam}$ and 
$\bnu_{\Lam^0}$-a.a. $\bOm^{*0}\in W^{\,*}_{\Lam^0}$.
\vskip .5 truecm

{\bf Remark 3.1.} The measure $\bmu_\Gam$ inherits the property from
Remark 2.3 and is concentrated on the subset in $W^*_\Gam$ formed by 
(infinite) loop configurations $\bOm^*_\Gam$ such that $\forall$ $\tau\in [0,\beta]$,
the section $\bOm^*_\Lam (\tau)$ has $\leq\kappa$ particles at each vertex $i\in\Lam$. 
\vskip .5 truecm

Given a measure $\bmu\in\bfG$, we associate with it a normalized linear functional 
$\bvphi =\bvphi_{\bmu}$ on  
the quasilocal C$^*$-algebra $\bfB$. First, we set 
$$\begin{array}{l}\diy
{\bF}^{\Lam^0}(\bx^{*0},\by^{*0})=
\int_{\ocW^{\,*}_{\bx^0,\by^0}}\obbP^{\,*}_{\bx^0,\by^0}
(\rd \obOm^{*0})\oB (\obOm^{*0}){\mathbf 1}\left(\obOm^{*0}\in\cF^{\Lam^0}\right)\\
\diy\qquad\times\int_{W^{\,*}_{\Gam\setminus\Lam}}
\bmu^{\Gam\setminus\Lam}\left(\rd\bOm^*_{\Gam\setminus\Lam}\right)
{\mathbf 1}(\bOm^*_{\Gam\setminus\Lam}\in
\cF^{\Lam^0}) 
{\wh q}^{\Lam^0}_{\Gam\setminus\Lam}\left(\obOm^{*0}\big|
\bOm^*_{\Gam\setminus\Lam}\right) \end{array}\eqno (3.1.5)$$
where
$$\begin{array}{r}\diy
{\wh q}^{\Lam^0}_{\Gam\setminus\Lam}\left(\obOm^{*0}\big|\bOm^*_{\Gam\setminus\Lam}\right) 
=\exp\,\left[-{\bh}^{\Lam^0}\left(\obOm^{*0}\big|\bOm^*_{\Gam\setminus\Lam}\right)\right]\qquad{}\\
\diy\times\frac{{\wh\bXi}^{\Lam^0}_\Gam \left( \Lam\setminus\Lam^0\big|\obOm^{*0}\vee\bOm^*_{\Gam\setminus\Lam}\right)}{\bXi_\Gam \left( \Lam\big|\bOm^*_{\Gam\setminus\Lam}\right)}.\end{array}\eqno (3.1.6)$$

This defines a kernel ${\bF}^{\Lam^0}(\bx^{*0},\by^{*0})$, $\bx^{*0},
\by^{*0}\in M^{*\Lam^0}$, where $\Lam^0\subset\Gam$ is a finite set of sites.
It is worth reminding the reader of the presence of the indicator functionals  
${\mathbf 1}\left(\;\cdot\;\in\cF^{\Lam^0}\right)$ in (3.1.5) and (3.1.6) (in the
integral for ${\wh\bXi}^{\Lam^0}_\Gam \left( \Lam\setminus\Lam^0\big|\obOm^{*0}\vee\bOm^*_{\Gam\setminus\Lam}\right)$). These 
indicators guarantee the compatibility property:  $\forall$ finite $\Lam^0\subset\Lam^1$, 
$$\begin{array}{l}{\bF}^{\Lam^0}(\bx^{*0},\by^{*0})\\
\diy\qquad =\int_{M^{*\Lam^1\setminus\Lam^0}}
\prod\limits_{j\in\Lam^1\setminus\Lam^0}\prod\limits_{z\in\bz^*(j)}v({\rd}z) 
{\bF}^{\Lam^1}(\bx^{*0}\vee\bz^*_{\Lam^1\setminus\Lam^0},
\by^{*0}\vee\bz^*_{\Lam^1\setminus\Lam^0}).\end{array}\eqno (3.1.7)$$

Next, we identify the operator $\bR^{\Lam^0}$ (a candidate for the RDM in volume
$\Lam^0$) as an integral operator acting
in $\bcH_{\Lam^0}$ by
$$\left(\bR^{\Lam^0}\bphi\right)(\bx^{*0})=\int_{M^{*\Lam^0}}
{\bF}^{\Lam^0}(\bx^{*0},\by^{*0})\bphi (\by^{*0})\rd\by^{*0}.\eqno (3.1.8)$$
Eqn (3.1.7) implies that
$${\rtr}_{\bcH_{\Lam^1\setminus\Lam^0}}\bR^{\Lam^1}=\bR^{\Lam^0}.$$ 
\vskip .5 truecm

{\bf Definition 3.3.} 
The functional $\bvphi\in\bfG$ is identified with the (compatible) 
family of operators $\bR^{\Lam^0}$. If the operators $\bR^{\Lam^0}$ are 
positive definite (a property that is not claimed to be automatically fulfilled), we 
again call it an FK-DLR state in the 
infinite volume (for given values of activity $z$ and inverse temperature 
$\beta$). To stress the dependence on $z$ and $\beta$, we sometimes employ
the notation $\bfG (z,\beta )$. $\quad\Diamond$ 
\vskip .5 truecm

{\bf 3.2. Theorems on existence and properties of FK-DLR states.}
We are now in position to state results about class $\bfG$. We assume the 
conditions on the potentials $U^{(1)}$ and $U^{(2)}$ from the previous 
section, including the hard-core condition for $U^{(1)}$. 
\vskip 0.5 truecm

{\bf Theorem 3.1.} {\sl For all $z,\beta\in (0,+\infty )$ satisfying
{\rm{(1.3.9)}}, any 
limiting Gibbs state $\bvphi\in\bfG^0$ (see Theorem {\rm{1.1}}) lies in 
$\bfG$. Therefore, the class of states $\bfG$ is non-empty. $\quad\lhd$}
\vskip .5 truecm

{\bf Theorem 3.2.} {\sl Under condistion {\rm{(1.3.9)}}, any 
FK-DLR state $\bvphi\in\bfG$ is ${\ttG}$-invariant, in the sense that, 
$\forall$ finite $\Lam^0\subset\Gam$ and $\forall$ ${\ttg}\in{\ttG}$, 
the RDM $\bR^{\Lam^0}$ satisfies {\rm{(1.5.4)}}. Consequently, Eqn 
{\rm{(1.5.5)}} holds true.$\quad\lhd$}
\vskip 2 truecm

\centerline{\bf 4. Proof of Theorems 1.1, 1.2 and 3.1.--3.3}
\vskip 1 truecm

{\bf 4.1. Proof of Theorems 1.1 and 3.1.} The proof is based on 
the same 
approach as that used in \cite{KS1}. First, given $\Lam^0\subset\Gam$,
we establish compactness of the sequence of the RDMKs 
${\bF}^{\Lam^0}_\Lam (\bx^{*0},\by^{*0})$ and ${\bF}^{\Lam^0}_{\Lam|\obx^*_{{\oGam}\setminus\Lam}}(\bx^{*0},\by^{*0})$ 
(see (2.3.2)--(2.3.7)) as functions of variables
$\bx^{*0}=\{\bx^{*0}(i)\},\by^{*0}=\{\by^{*0}(i)\}\in M^{*\Lam^0}$, with 
$$\sharp\,\bx^{*0}_\Lam
=\sharp\,\by^{*0}_\Lam\;\hbox{ and }\; 
\sharp\,\bx^{*0}(i),\sharp\,\by^{*0}(i)<\kappa,\;\;i\in\Lam ,$$
when $\Lam\nearrow\Gam$. 
Then we use Lemma 1.1 
from \cite{KS1} to derive that the sequence of the RDMs $\bR^{\Lam^0}_{\Lam}$
and $\bR^{\Lam^0}_{\Lam|\obx^*_{{\oGam}\setminus\Lam}}$ is compact in the trace-norm
operator topology in $\cH_{\Lam^0}$. 
 
To verify compactness of the RDMKs ${\bF}^{\Lam^0}_\Lam (\bx^{*0},\by^{*0})$ and ${\bF}^{\Lam^0}_{\Lam|\obx^*_{{\oGam}\setminus\Lam}}(\bx^{*0},\by^{*0})$ we, again as in \cite{KS1},
use the Ascoli--Arzela theorem, which requires the properties of uniform boundedness and
equicontinuity. These properties follow from 
\vskip .5 truecm

{\bf Lemma 4.1.} (i) {\sl Under condition (1.3.9) the RDMKs 
${\bF}^{\Lam^0}_\Lam (\bx^{*0},\by^{*0})$ and ${\bF}^{\Lam^0}_{\Lam|\obx^*_{{\oGam}\setminus\Lam}}(\bx^{*0},\by^{*0})$ admit the 
bounds
$${\bF}^{\Lam^0}_\Lam (\bx^{*0},\by^{*0}),\,
{\bF}^{\Lam^0}_{\Lam|\obx^*_{{\oGam}\setminus\Lam}}(\bx^{*0},\by^{*0})\\
\quad\leq\big[(\kappa\sharp\Lam^0)!\big]({\wh p}_M)^{\kappa\sharp\Lam^0}
\Phi^{\sharp\Lam^0}\eqno (4.1.1)$$
where
$$\Phi =\sum\limits_{k\geq 1}z^k\exp\,(k\Theta ),\;\hbox{ with }\;\Theta
=\kappa\beta \big({\ov U}^{(1)}
+\kappa{\ov U}^{(2)}+\kappa{\oJ}(1){\ov V}\big).
\eqno (4.1.2)$$ 
(Note that $\Phi <\infty$ under the assumption {\rm{(1.3.9)}}.) Let
$p^{(k)}_M$ yields the supremum of the transition function over time 
$k\beta$ for Brownian motion 
on the torus $M$:
$$p^{(k)}_M={\operatornamewithlimits{\sup}\limits_{x,y\in M}} p^{k\beta}(x,y)=
p^{k\beta}(0,0),$$
and ${\wh p}_M={\operatornamewithlimits{\sup}\limits_{k\geq 1}} p^{(k)}_M$. 
Finally, the upper-bound values ${\ov U}^{(1)}$, ${\ov U}^{(2)}$, ${\oJ}(1)$ and ${\ov V}$
have been determined in {\rm{(1.3.1), (1.3.2)}} and {\rm{(1.3.5)}}.} 
 
(ii) {\sl The gradients of the RDMKs ${\bF}^{\Lam^0}_\Lam (\bx^{*0},\by^{*0})$ and ${\bF}^{\Lam^0}_{\Lam|\obx^*_{{\oGam}\setminus\Lam}}(\bx^{*0},\by^{*0})$ satisfy:  
$\forall$ $i\in\Lam$ and $x\in\bx^{*0}(i)$, $y\in\by^{*0}(i)$,
$$\begin{array}{l}\left|\nabla_{x}
{\bF}^{\Lam^0}_\Lam(\bx^{*0},\by^{*0})\right|,\,\left|
\nabla_{x}{\bF}^{\Lam^0}_{\Lam|\obx^*_{{\oGam}\setminus\Lam}}
(\bx^0,\by^0)\right|\\
\quad\left|\nabla_{y}
{\bF}^{\Lam^0}_\Lam (\bx^{*0},\by^{*0})\right|,\,\left|
\nabla_{y}{\bF}^{\Lam^0}_{\Lam|\obx^*_{{\oGam}\setminus\Lam}}
(\bx^0,\by^0)\right|\\
\qquad\leq (\sharp\Lam^0)\Theta\Phi' 
{\wh p}_M \big[(\kappa\sharp\Lam^0)!\big]({\wh p}_M)^{\kappa\sharp\Lam^0}
\Phi^{\sharp\Lam^0}\end{array}\eqno (4.1.3)$$
where 
$$\Phi'=\sum\limits_{k\geq 1}kz^k\exp\,(k\Theta ).
\eqno (4.1.4)$$
(Again, $\Phi'<\infty$ under the condition {\rm{(1.3.9)}}.)
The ingredients ${\wh p}_M$ 
and ${\ov U}^{(1)}$, ${\ov U}^{(2)}$, ${\oJ}(1)$ and ${\ov V}$ as in 
statement} (i).$\lhd$ 
\vskip .5 truecm

{\it Proof of Lemma} 4.1. First, observe that ${\wh p}_M<\infty$ on a 
compact manifold. Bound (4.1.2) is established in a direct fashion.
First, we  majorize the energy 
$${\bh}^{\Lam^0} (\obOm^{*0}|\bOm^*_{\Lam\setminus\Lam^0})
={\bh}^{\Lam^0}(\obOm^{*0}) 
+\bh(\bOm^*_{\Lam \setminus\Lam^0}|\,|\obOm^{*0})$$
contributing to the RHS in (2.3.4) and (2.3.5) and the energy
$${\bh}^{\Lam^0}(\obOm^{*0}|\bOm^*_{\Lam\setminus\Lam^0}\vee
\obx^*_{{\oGam}\setminus\Lam})
={\bh}^{\Lam^0}(\obOm^{*0}|\obx^*_{{\oGam}\setminus\Lam})+
\bh(\bOm^*_{\Lam \setminus\Lam^0}|\,|\obOm^{*0})$$ 
contributing to the RHS in (2.3.6) and (2.3.7). This yields 
the factor
$$\prod\limits_{\oom\in\obOm^{*0}}z^{k(\oom^*)}\exp\,\big[
k(\oom^*)\Theta\big].$$

Next, we majorize the integral $\diy =\int_{\ocW^{\,*}_{\bx^{*0},\by^{*0}}}
\obbP^{\,*}_{\bx^{*0},\by^{*0}}({\rd}\obOm^{*0})$
in (2.3.4) and (2.3.6); this gives the factor 
$$\begin{array}{l}\diy (\sharp\Lam^0) {\wh p}_M \big[(\kappa\sharp\Lam^0)!\big]
({\wh p}_M)^{\kappa\sharp\Lam^0}\Phi^{\sharp\Lam^0}.
\end{array}$$ 
The aftermath are the ratios (2.3.2) and (2.3.3) with $\bx^{*0}=\by^{*0}=\emptyset$;
they do not exceed $1$. 

Passing to (4.1.3), let us discuss the gradients $\nabla_x$ only.  
(The gradients in the entries of $\nabla_y$ are included by symmetry.)
The gradient in (4.1.3), of course, affects only the numerators
${\wh\bXi}^{\Lam^0}_\Lam (\bx^{*0},\by^{*0};\Lam\setminus\Lam^0)$ and
${\wh\bXi}^{\Lam^0}_\Lam(\bx^{*0},\by^{*0};\Lam\setminus
\Lam^0\,|\,\obx^*_{{\oGam}\setminus\Lam})$ in (2.3.2) and (2.3.3).
The bounds (4.1.3) are done essentially as in \cite{KS1}. For definiteness, we 
discuss the case of the RDMK ${\bF}^{\Lam^0}_\Lam (\bx^{*0},\by^{*0})$;
the RDMK ${\bF}^{\Lam^0}_{\Lam|\obx^*_{{\oGam}\setminus\Lam}}
(\bx^0,\by^0)$ is treated similarly. There are two 
contributions into the gradient: one comes from varying 
the measure $\obbP^{\,*}_{\bx^{*0},\by^{*0}}({\rd}\obOm^{*0})$, 
the other from
varying the functional 
$\exp\left[-{\bh}^{\Lam^0}(\obOm^{*0})-{\bh}^{\Lam\setminus\Lam^0} (\bOm^*_{\Lam \setminus\Lam^0}|\obOm^{*0})\right]$. 

The first
contribution can again be uniformly bounded in terms of the constant ${\wh p}_M$. 
The detailed argument, as in \cite{KS1}, includes
a deformation of a trajectory and is done similarly to \cite{KS1} (the presence
of jumps does not change the argument because ${\wh p}_M$ yields a uniform
bound in (2.1.7)). 
 
The second contribution yields, again as in \cite{KS1}, an expression of the form
$$\begin{array}{l}\diy\int_{\oW^{\,*}_{\bx^{*0},\by^{*0}}}
\obbP^{\,*}_{\bx^{*0},\by^{*0}}({\rd}\obOm^{*0})\sum_{i\in\Lam^0}\sum_{x\in\bx^{*0}(i)}
{\wt\bh}_{x,i}(\oOm^*_{(x,i),\gam^0(x,i)},\bOm^*_{\Lam\setminus\Lam^0})\\    
\diy\qquad\times\exp\,\left[-{\bh}^{\Lam^0}(\obOm^{*0})-{\bh}^{\Lam\setminus\Lam^0}
(\bOm^*_{\Lam\setminus\Lam^0}|\obOm^{*0})\right]\end{array}\eqno (4.1.5)$$
where the functional ${\wt\bh}_{x,i}(\oOm^*_{(x,i),\gam^0(x,i)},
\bOm^*_{\Lam\setminus\Lam^0})$
is uniformly bounded. Combining this an upper bound similar to (4.1.2)
yields the desired estimate for the gradients in (4.1.3). $\quad\Box$
\vskip .5 truecm

Hence, we can guarantee that the RDMs $\bR^{\Lam^0}_{\Lam}$
and $\bR^{\Lam^0}_{\Lam|\obx^*_{{\oGam}\setminus\Lam}}$ converge to a limiting 
RDM $\bR^{\Lam^0}$ along a subsequence in $\Lam\nearrow\Gam$. The diagonal process
yields convergence for every finite $\Lam^0\subset\Gam$. A parallel argument leads 
to compactness of the measures $\mu_\Lam^{\Lam^0}$ for any given $\Lam^0$ as
$\Lam\nearrow\Gam$. We only give here a sketch of the corresponding argument, 
stressing differences with its counterpart in \cite{KS1}.  

In the probabilistic terminology, measures $\mu_\Lam$ represent random marked 
point fields on $M\times \Gam$ with marks from the space $W^*=W^*_{0,0}$ where
$\diy W^*_0=\;{\operatornamewithlimits{\bigcup}\limits_{k\geq 1}}W^{k\beta}_0$
and $W^{k\beta}_0$ is the space of loops of time-length $k\beta$ starting 
and finishing at $0\in M$ and exhibiting jumps, i.e., changes of the index.
(The space $W^*_{x,i}$ introduced in Definitions 2.1 and 2.5 can be considered 
as a copy of $W^*$ placed at site $i\in\Gam$ and point $x\in M$.) The measure 
$\mu_\Lam^{\Lam^0}$ describes the restriction of $\mu_\Lam$ to volume 
$\Lam^0$ (i.e., to the sigma-algebra $\fW^*_{\Lam^0}$) and is given by
its Radon-Nikodym derivative $p_\Lam^{\Lam^0}$ relative to the reference 
measure ${\rd}\bOm^*_{\Lam^0}$ on $W^*_{\Lam^0}$ (cf. (2.2.11), (2.4.3)). 
The reference measure is sigma-finite. Moreover, under the condition (1.3.9), 
the value $p_\Lam^{\Lam^0}(\bOm^*_{\Lam^0})$ is uniformly bounded (in both 
$\Lam\nearrow\Gam$ and $\bOm^*_{\Lam^0}\in W^*_{\Lam^0}$). This enables us 
to verify tightness
of the family of measures $\{\mu_\Lam^{\Lam^0},\,\Lam\nearrow\Gam\}$ and
apply the Prokhorov theorem. Next, we use the compatibility property     
of the limit-point measures $\mu_\Gam^{\Lam^0}$ and apply the Kolmogorov theorem.
This establishes the existence of the limit-point measure $\mu_\Gam$.  

By construction, and owing
to Lemmas 2.4 and 2.5, the limiting
family $\{\bR^{\Lam^0}\}$ yields a state belonging to the class $\bfG$. This completes
the proof of Theorems 1.1 and 3.1.$\quad\Box$
\vskip 1 truecm

{\bf 4.2. Proof of Theorems 1.2 and 3.2.} The assertion of Theorem 1.2 is included
in Theorem 3.2. Therefore, we will focus on the proof of the latter. The proof 
based on the analysis of the conditional
RDMFs $q^{\Lam^0}_{\Gam\setminus\Lam}
\left(\obOm^{*0}\big|\bOm^*_{\Gam\setminus\Lam}\right)$
and ${\wh q}^{\Lam^0}_{\Gam\setminus\Lam}
\left(\obOm^{*0}\big|\bOm^*_{\Gam\setminus\Lam}\right)$ introduced in Eqn (3.1.4)
and (3.1.5). For definiteness, we assume that vertex $o\in\Lam^0$, so that 
$\Lam^0$ lies in the ball $\Lam_n$ for $n$ large enough. As in \cite{KS1}, the problem 
is reduced to checking that 
$\forall$ $z,\beta\in (0,\infty)$ satisfying (1.3.9), ${\ttg}\in{\ttG}$ and finite $\Lam^0\subset\Gam$, 
$$\lim_{n\to\infty}\frac{
{\wh q}^{\Lam^0}_{\Gam\setminus\Lam (n)}\left({\ttg}\obOm^{*0}\big|\bOm^*_{\Gam\setminus\Lam (n)}\right)}{
{\wh q}^{\Lam^0}_{\Gam\setminus\Lam (n)}\left(\obOm^{*0}\big|\bOm^*_{\Gam\setminus\Lam (n)}\right)}=1;
\eqno (4.2.1)$$
here we need to establish this convergence (4.2.1) uniformly in the argument $\bOm^*_{\Gam\setminus\Lam (n)}=\{\bOm^*(i),i\in\Gam\setminus\Lam (n)\}$ with
$\sharp\,\bOm^*(i)\leq\kappa$ and in $\obOm^{*0}$ outside a set of the $
\obbP^{*\beta}_{\bx^0,\by^0}$-measure tending to $0$ as $n\to\infty$. The latter is 
formed by path configurations $\obOm^{*0}$ that contain trajectories visiting sites $i\in\Gam\setminus
\Lam ({\ovr}(n))$ where ${\ovr}(n)$ grows with $n$; see Lemma 4.2 below. 
The action of ${\ttg}$ upon a path configuration
$\obOm^{*0}=\{\oOm^{\,*}_{(x,i),\gam^0(x,i)},i\in\Lam^0,x\in\bx^{*0}\}$ is defined by
$${\ttg}\obOm^{*0}=\{{\ttg}\oOm^*_{(x,i),\gam^0(x,i)}\}\;\hbox{ where }\;
\left({\ttg}\oOm^*_{(x,i),\gam^0(x,i)}\right)(\tau)
={\ttg}\left(\oOm^*_{(x,i),\gam^0(x,i)}(\tau)\right).$$

We want to establish that   
$\forall$ $a \in (1,\infty )$, for any $n$ large enough, the conditional RDMFs
satisfy
$$\begin{array}{r}a{\wh q}^{\Lam^0}_{\Gam\setminus\Lam (n)}({\ttg}\obOm^{*0}|
\bOm^*_{\Gam\setminus\Lam (n)})
+a{\wh q}^{\Lam^0}_{\Gam\setminus\Lam (n)}({\ttg}^{-1}\obOm^{*0}|
\bOm^*_{\Gam\setminus\Lam (n)})\qquad{}\\
\geq 2{\wh q}^{\Lam^0}_{\Gam\setminus\Lam (n)}(\obOm^{*0}|
\bOm^*_{\Gam\setminus\Lam (n)}).\end{array}\eqno (4.2.2)$$

As in \cite{KS1}, we deduce (4.2.2) with the help of a special construction
of `tuned' actions $\bttg_{\Lam (n)\setminus\Lam^0}$
on loop configurations $\bom_{\Lam (n)\setminus\Lam^0}$
(over which there is integration performed in the numerators \\
${\wh\bXi}^{\Lam^0}_\Gam \left( \Lam (n)\setminus\Lam^0\big|{\ttg}\obOm^{*0}
\vee\bOm^*_{\Gam\setminus\Lam (n)}\right)$ and 
${\wh\bXi}^{\Lam^0}_\Gam \left( \Lam (n)\setminus\Lam^0\big|{\ttg}^{-1}\obOm^{*0}
\vee\bOm^*_{\Gam\setminus\Lam (n)}\right)$ \\
in the expression for ${\wh q}^{\Lam^0}_{\Gam\setminus\Lam (n)}({\ttg}\obOm^{*0}|
\bOm^*_{\Gam\setminus\Lam (n)})$ and $
{\wh q}^{\Lam^0}_{\Gam\setminus\Lam (n)}({\ttg}^{-1}\obOm^{*0}|
\bOm^*_{\Gam\setminus\Lam (n)})$). \\
The tuning in $\bttg_{\Lam (n)\setminus\Lam^0}$ is chosen so that 
it approaches ${\tt e}$ (or the $d$-dimensional zero vector in 
the additive form of writing), 
the neutral element of ${\ttG}$, while we move from $\Lam^0$ towards $\Gam
\setminus\Lam (n)$. 

Formally, (4.2.2) follows from the estimate (4.2.3) below: $\forall$ finite 
$\Lam^0\subset\Gam$, 
$\obOm^{*0}\in\oW_\Lam (n)$, ${\ttg}\in{\ttG}$ and $a \in (1,\infty )$,
for any $n$ large enough, $\forall$ $\bOm^*_{\Lam (n)\setminus\Lam^0}=\{\bOm^*(i),
i\in\Lam (n)\setminus\Lam^0\}$ and 
$\bOm^*_{\Gam\setminus\Lam (n)}=\{\bOm^*(i),
i\in\Gam\setminus\Lam (n)\}$ with $\sharp\,\bOm^*(i)\leq\kappa$, 
$$\begin{array}{l}\diy\frac{a}{2}\exp\,\Big[-{\bh}^{\Lam (n)}(({\ttg}\obOm^{*0})\vee
(\bttg_{\Lam (n)\setminus\Lam^0}\bOm^*_{\Lam (n)\setminus\Lam^0})|
\bOm^*_{\Gam\setminus\Lam (n)})\Big]\\
\quad\diy +
\frac{a}{2}\exp\,\Big[-{\bh}^{\Lam (n)}(({\ttg}^{-1}\obOm^{*0})\vee
(\bttg^{-1}_{\Lam (n)\setminus\Lam^0}\bOm^*_{\Lam (n)\setminus\Lam^0})|
\bOm^*_{\Gam\setminus\Lam (n)})\Big]\\
\qquad\quad\geq\diy
\exp\,\Big[-{\bh}^{\Lam (n)}(\obOm^{*0}\vee\bOm^*_{\Lam (n)\setminus\Lam^0}|
\bOm^*_{\Gam\setminus\Lam (n)})\Big]\,.
\end{array}\eqno (4.2.3)$$

In (4.2.3), the loop configuration 
$\bttg_{\Lam (n)\setminus\Lam^0}\bOm^*_{\Lam (n)\setminus\Lam^0}$
is determined by specifying its temporal section 
$\{\bttg_{\Lam (n)\setminus\Lam^0}\Om^*_{x,i}(\tau +\beta m), i\in\Lam (n)\setminus\Lam^0, x\in\bx^*(i),0\leq m<k_{x,i}\}$. That is, we need to specify the 
sections 
$$\big(u(\tau +\beta m;\bttg_{\Lam (n)\setminus\Lam^0}\Om^*_{x,i}),
l(\tau +\beta m;\bttg_{\Lam (n)\setminus\Lam^0}\Om^*_{x,i})\big)$$  
for loops $\Om^*_{x,i}$ constituting $\bttg_{\Lam (n)\setminus\Lam^0}\bOm^*_{\Lam (n)\setminus\Lam^0}$. To this end we set:
$$\begin{array}{l}
u(\tau +\beta m;\bttg_{\Lam (n)\setminus\Lam^0}\Om^*_{x,i})=
{\ttg}^{(n)}_ju(\tau +\beta m; \Om^*_{x,i})\\
\qquad\qquad\hbox{ if }\;l(\tau +\beta m;\bttg_{\Lam (n)\setminus\Lam^0}\Om^*_{x,i})=j.
\end{array}\eqno (4.2.4)$$
In other words, we apply the action ${\ttg}^{(n)}_j$ to the temporal sections of all 
loops $\Om^*_{x,i}$ located at vertex $j$ at a given time, regardless of 
position of their initial points $(x,i)$ in $\Lam (n)\setminus\Lam^0$. 

Observe that (4.2.2) is deduced from (4.2.3) by integrating in ${\rd}
\bOm^*_{\Lam (n)\setminus\Lam^0}$ and normalizing
by $\bXi_{\Lam (n)\setminus\Lam^0}(\bOm^*_{\Gam\setminus\Lam (n)})$;
cf. Eqn (2.3.4) with $\Lam'=\Lam (n)$.
(The Jacobian of the map $\bOm^*_{\Lam (n)
\setminus\Lam^0}\mapsto\bttg_{\Lam (n)\setminus\Lam^0}
\bOm^*_{\Lam (n)\setminus\Lam^0}$ equals $1$.)

Thus, our aim becomes to prove (4.2.3). The tuned family 
$\bttg_{\Lam (n)\setminus\Lam^0}$ is composed of individual
actions ${\ttg}^{(n)}_j\in{\ttG}$: 
$$\bttg_{\Lam (n)\setminus\Lam^0}=\{{\ttg}^{(n)}_j,\;j\in\Lam (n)
\setminus\Lam^0\}.\eqno (4.2.5)$$
Elements ${\ttg}^{(n)}_j$ are powers (multiples, in the additive parlance) of element 
${\ttg}\in{\ttG}$ figuring in (4.2.1)--(4.2.3) (respectively, of the corresponding 
vector $\utheta\in M$; cf. Eqn 
(1.2.1)) and defined as follows. Let $\utheta^{(n)}_j$ denote
the vector from $M$ corresponding to ${\ttg}^{(n)}_j$, we select  
positive integer values $\ovr (n)=\big\lceil\log\;(1+n)\big\rceil$ and set:
$$\utheta^{(n)}_j=\utheta\,\upsilon (n,j)\eqno (4.2.6)$$
where 
$$\upsilon (n,j)
=\begin{cases}1,&{\tt d}(o,j)\leq\ovr (n),\\
\vartheta\big({\tt d}(j,o)-\ovr (n),n-\ovr (n)\big),&{\tt d}(o,j)>\ovr (n)
.\end{cases}\eqno (4.2.7)$$
In turn, the function $\vartheta$ is chosen to satisfy
$$\vartheta (a,b)={\mathbf 1}(a\leq 0)+
\frac{{\mathbf 1}(0<a<b)}{Q(b)}\int_a^bz(u){\rd}u,\;\;
a,b\in\bbR ,\eqno (4.2.8)$$
with
$$\begin{array}{l}\diy Q(b)=\int_0^b \zeta (u){\rd}u\sim\,\log\log\,b,\\
\qquad\diy\hbox{where}
\;\zeta (u)={\mathbf 1}(u\leq 2)+{\mathbf 1}(u>2)
\frac{1}{u\ln\,u},\;b>0.\end{array}\Box\eqno (4.2.9)$$
\vskip 1 truecm

{\bf Lemma 4.2.} {\sl Given $z,\beta\in (0,\infty )$ satisfying (1.3.9) and a finite set 
$\Lam^0$, there exists a constant $C\in (0,\infty )$ such that $\forall$ $\bx^0,\by^0\in M^{*\Lam^0}$
the set of path configurations $\obOm^{*0}\in
\oW^{\,*}_{\bx^{*0},\by^{*0}}$ with ${\bh}^{\Lam^0}(\obOm^{*0})<+\infty$ 
which include trajectories visiting points in
$\Gam\setminus\Lam ({\ovr}(n))$ has the $
\obbP^{\,*}_{\bx^0,\by^0}$-measure that does not exceed 
$\diy\frac{C}{\left(\big\lceil (1+r(n))\big\rceil\right)!}$.$\quad\lhd$}
\vskip .5 truecm

{\it Proof of Lemma} 4.2. The condition that ${\bh}^{\Lam^0}(\obOm^{*0})<+\infty$ implies that
the total number of sub-trajectories of length $\beta$ in $\obOm^{*0}$ does not exceed
$\kappa\times\sharp\Lam^0$ which is a fixed value in the context of the lemma. Each such trajectory
has a Poisson number of jumps; this produces the factor $1/\left(\big\lceil (1+r(n))\big\rceil\right)!$.
$\quad\Box$.  
\vskip 1 truecm

Back to the proof Theorem 3.2: let $\bttg_{\Lam (n)\setminus\Lam^0}^{-1}$ be the 
collection of the inverse elements:
$$\bttg_{\Lam (n)\setminus\Lam^0}^{-1}=
\left\{{{\ttg}_j^{(n)}}^{-1},\;j\in\Lam (n)\setminus\Lam^0\right\}.$$ 
The vectors corresponding to ${{\ttg}_j^{(n)}}^{-1}$ are $-\utheta^{(n)}_j\in M$.
We will use this specification for ${\ttg}^{(n)}_j$ and ${{\ttg}_j^{(n)}}^{-1}$ for 
$j\in\Lam (n)$, or even for $j\in\Gam$, as it agrees with the requirement 
that ${\ttg}^{(n)}_j\equiv{\ttg}$
when $j\in\Lam^0$ and ${\ttg}^{(n)}_j\equiv{\tt e}$ for $j\in\Gam
\setminus\Lam (n)$. Accordingly, we will use the notation $\bttg_{\Lam (n)}
=\{{\ttg}^{(n)}_j,j\in\Lam (n)\}$.
 
Observe that the tuned family 
$\bttg_{\Lam (n)\setminus\Lam^0}$ does not change the contribution into the energy 
functional ${\bh}^{\Lam (n)|\Gam
\setminus\Lam (n)}$ coming from potentials $U^{(`1)}$ and $U^{(2)}$: it affects only 
contributions from potential $V$.
  
The Taylor formula for function $V$, together with the above identification
of vectors $\utheta^{(n)}_j$, gives:
$$\begin{array}{l}
\Big|V\left({\ttg}^{(n)}_jx,{\ttg}^{(n)}_{j'}x'\right)
+V\left({{\ttg}^{(n)}_j}^{-1}x,
{{\ttg}^{(n)}_{j'}}^{-1}x'\right)-2V(x,x')\Big|\\ \;\\
\qquad\qquad\leq C\,
|\utheta |^2\left|\upsilon (n,j)-\upsilon (n,j')\right|^2
{\ov V}\,,x,x'\in M.\end{array}\eqno (4.2.9)$$ 
Here $C\in (0,\infty )$ is a constant, $|\utheta |$
stands for the norm of the vector $\utheta$ representing 
the element ${\ttg}$ and the value ${\ov V}$ is taken
from (1.3.1).

Next, the square $\left|\upsilon (n,j)-\upsilon (n,j')\right|^2$
can be specified as 
$$\begin{array}{l}
\left|\upsilon (n,j)-\upsilon (n,j')\right|^2\\
\diy\quad =\begin{cases}0,
\;\hbox{ if }\;{\tt d}(j,o),
{\tt d}(j',o)\leq \ovr (n),\\
0,\;\hbox{ if }\;{\tt d}(j,o),{\tt d}(j',o)\geq n,\\
\big[\vartheta ({\tt d}(j,o)-\ovr (n),n-\ovr (n))\\
\quad -\vartheta ({\tt d}(j',o)-\ovr (n),n-\ovr (n))\big]^2,\\
\qquad\hbox{ if }\;\ovr (n)<{\tt d}(j,o),{\tt d}(j',o)\leq n,\\
\vartheta ({\tt d}(j,o)-\ovr (n),n-\ovr (n))^2,\\
\qquad\hbox{ if }\;\ovr (n)<{\tt d}(j,o)\leq n,
{\tt d}(j',o)\in ]\ovr (n),n[,\\
\vartheta ({\tt d}(j',o)-\ovr (n),n-\ovr (n))^2,\\
\qquad\hbox{ if }\;\ovr (n)<{\tt d}(j',o)\leq n,
{\tt d}(j,o)\in ]\ovr(n),n[.\end{cases}\end{array}\eqno (4.2.11)$$

By using convexity of the function exp and Eqn (4.2.10), $\forall$ $a>1$,
$$\begin{array}{l}\diy\frac{a}{2}\exp\,\Big[- 
{\bh}^{\Lam (n)}\Big({\bttg}_{\Lam (n)}\big(\obOm^{*0}\vee
\bOm^*_{\Lam (n)\setminus\Lam^0}\big)\big|
\bOm^*_{\Gam\setminus\Lam (n)}\Big)\Big]\\
\quad +\diy\frac{a}{2}\exp\,\Big[- 
{\bh}^{\Lam (n)}\Big({\bttg}_{\Lam (n)}^{-1}\big(
\obOm^{*0}\vee
\bOm^*_{\Lam (n)\setminus\Lam^0}\big)\big| 
\bOm^*_{\Gam\setminus\Lam (n)}\Big)\Big]\\
\;\;\diy\geq a\exp\;\bigg[-\frac{1}{2}
{\bh}^{\Lam (n)|\Gam\setminus\Lam (n)}\Big({\bttg}_{\Lam (n)}\big(
\obOm^{*0}\vee
\bOm^*_{\Lam (n)\setminus\Lam^0}\big), 
\bOm^*_{\Gam\setminus\Lam (n)}\Big)\\
\qquad -\diy\frac{1}{2}
{\bh}^{\Lam (n)}\Big({\bttg}_{\Lam (n)}^{-1}\big(
\obOm^{*0}\vee
\bOm^*_{\Lam (n)\setminus\Lam^0}\big)\big|
\bOm^*_{\Gam\setminus\Lam (n)}\Big)\bigg]\\
\;\;\geq a\exp\,\Big[
-{\bh}^{\Lam (n)}\Big(
\obOm^{*0}\vee\bOm^*_{\Lam (n)\setminus\Lam^0}\big|
\bOm^*_{\Gam\setminus\Lam (n)}\Big)\Big]
e^{-C\Upsilon /2}\end{array}\eqno (4.2.12)$$
where
$$\Upsilon =\Upsilon (n,{\ttg})=\beta\kappa^2\sum_{(j,j')\in\Lam (n)\times\Gam}
J({\ttd}(j,j'))\left|\upsilon (n,j)-\upsilon (n,j')\right|^2.\eqno (4.2.13)$$

The next observation is that 
$$\begin{array}{l}\Upsilon\leq 3\beta\kappa^2|\utheta |^2
\sum\limits_{(j,j')\in\Lam (n)\times\Gam}
{\mathbf 1}\big({\tt d}(j,o)\leq {\tt d}(j',0)
\big)J({\ttd}(j,j'))\\ \;\\
\qquad\times\Big[\vartheta ({\tt d}(j,o)-\ovr (n),n-\ovr (n)) -
\vartheta ({\tt d}(j',o)-\ovr (n),n-\ovr (n))\Big]^2\end{array}\eqno (4.2.14)
$$
where, owing to the triangle inequality, for all 
$j,j': {\tt d}(j,o)\leq {\tt d}(j',o)$
$$\begin{array}{r}
0\leq\vartheta ({\tt d}(j,o)-\ovr (n),n-\ovr (n)) -
\vartheta ({\tt d}(j',o)-\ovr (n),n-\ovr (n))\qquad{}\\
\leq {\tt d}(j,j')\diy\frac{\zeta ({\tt d}(j,o)-\ovr)}{Q(n-\ovr (n))}.
\end{array}\eqno (4.2.15)$$
This yields
$$\begin{array}{l}
\Upsilon\leq\diy\beta\kappa^2\frac{3|\utheta |^2}{Q(n-\ovr (n))^2}\sum\limits_{(j,j')\in\Lam (n)\times\Gam}
J({\ttd}(j,j')){\tt d}(j,j')^2\zeta ({\tt d}(j,0)-\ovr (n))^2\\
\quad\leq\diy\frac{3|\utheta |^2}{Q(n-\ovr (n))^2}
\Big[\sup_{j\in\Gam}\sum\limits_{j'
\in\Gam}J_{j,j'}{\tt d}(j,j')^2 \Big]\sum\limits_{j\in\Lam_{n+r_0}}
\zeta ({\tt d}(j,0)-\ovr (n))^2\end{array}$$
where function $\zeta$ is determined in (4.2.9).

Therefore, it remains to estimate the sum $\sum\limits_{j\in\Lam_{n+r_0}}
\zeta ({\tt d}(j,0)-\ovr (n))^2$. To this end, observe that 
$u\zeta (u)<1$ when $u\in (3,\infty )$. The next remark is that the
number of sites in the sphere $\Sigma_n$ grows linearly with $n$.
Consequently, 
$$\begin{array}{l}\sum\limits_{j\in\Lam_{n+r_0}}
\zeta ({\tt d}(j,o)-\ovr (n))^2=\sum\limits_{1\leq k\leq n+r_0}\zeta (k-\ovr (n))
\sum\limits_{j\in\Sigma_k}\zeta (k-\ovr (n))\\
\qquad\qquad\leq C_0\sum\limits_{1\leq k\leq n+r_0}\zeta(k-\ovr (n))
\leq C_1Q(n+r_0-\ovr (n))\end{array}$$
and
$$\Upsilon\leq\frac{C}{Q(n-\ovr (n))}\to\infty,\;\hbox{ as }\;n\to\infty.$$

Therefore, given $a>1$ for $n$ large enough,
the term $ae^{-C\Upsilon /2}$ in the RHS of (4.2.14)
becomes $>1$. Hence,
$$\begin{array}{l}\diy\frac{a}{2}\exp\,\Big[- 
{\bh}^{\Lam (n)}\Big({\bttg}_{\Lam (n)}\big(
\obOm^{*0}\vee
\bOm^*_{\Lam (n)\setminus\Lam^0}\big)
\big|\bOm^*_{\Gam\setminus\Lam (n)}\Big)\Big]\\
\quad +\diy\frac{a}{2}\exp\,\Big[- 
{\bh}^{\Lam (n)}\Big({\bttg}_{\Lam (n)}^{-1}\big(
\obOm^{*0}\vee\bOm^*_{\Lam (n)\setminus\Lam^0}\big)\big|
\bOm^*_{\Gam\setminus\Lam (n)}\Big)\Big]\\
\qquad\qquad\qquad
\;\;\geq\exp\,\Big[-{\bh}^{\Lam (n)}\Big(\obOm^{*0}\vee
\bOm^*_{\Lam (n)\setminus\Lam^0}\big|
\bOm^*_{\Gam\setminus\Lam (n)}\Big)\Big]\,
\end{array}\eqno (4.2.16)$$

Eqn (4.2.16) implies that  the quantity 
$$\begin{array}{r}
{q}^{\Lam^0|\Gam\setminus\Lam (n)}
(\obOm^{*0}|\bOm_{\Gam\setminus\Lam (n)})=
\diy\int_{W_{\Lam (n)\setminus\Lam^0}}
{\rd}\bOm^*_{\Lam (n)\setminus\Lam^0} 
\quad\qquad{}\\
\diy\times\frac{\exp\big[-{\bh}^{\Lam^0}
(\obOm^{*0}\vee\bOm^*_{\Lam (n)\setminus\Lam^0}\big|
\bOm^*_{\Gam\setminus\Lam^0})
\big]}{\Xi_{\Lam (n)}(
\bOm^*_{\Gam\setminus\Lam (n)})},\end{array}\eqno (4.2.17)$$
obeys
$$\begin{array}{r}
\lim\limits_{n\to\infty}\Big[{q}_{\beta}^{\Lam^0|\Gam\setminus\Lam (n)}
({\ttg}\obOm^{*0}|\bOm^*_{\Gam\setminus\Lam (n)})
+{q}_{\beta}^{\Lam^0|\Gam\setminus\Lam (n)}
({\ttg}^{-1}\obOm^{*0}|\bOm^*_{\Gam\setminus\Lam (n)})\Big]\qquad{}\\
\geq 2\lim\limits_{n\to\infty}
{q}_{\beta}^{\Lam^0|\Gam\setminus\Lam (n)}
(\obOm^{*0}|\bOm^*_{\Gam\setminus\Lam (n)})\end{array}\eqno (4.2.18)$$
uniformly in boundary condition $\bom_{\Gam\setminus\Lam (n)}$. 
Integrating (4.2.18) in\\ ${\rd}\mu_{\Gam}^{\Gam\setminus\Lam (n)}
(\bom_{\Gam\setminus\Lam (n)})$ yields (4.2.3). $\quad\Box$
\vskip 1 truecm

{\bf Acknowledgement.}   
This work has been conducted under Grant\\ 2011/20133-0 provided by 
the FAPESP, Grant 2011.5.764.45.0 provided by The Reitoria of the 
Universidade de S\~{a}o Paulo and Grant 2012/04372-7
provided by the FAPESP. The authors
express their gratitude to NUMEC and IME, Universidade de S\~{a}o Paulo,
Brazil, for the warm hospitality.

\end{document}